\documentclass[authoryear,3p,11pt]{elsarticle}

\usepackage{amsmath,amssymb,mathtools,mathrsfs}

\usepackage{ulem}
\usepackage{cancel}

\usepackage[usenames,dvipsnames]{xcolor}

\newcommand{\beq}{\begin{equation}}
\newcommand{\eeq}{\end{equation}}

\newcommand{\upd}{\mathrm{d}}
\newcommand{\mudot}{\dot{\mu}}
\newcommand{\A}{\mathcal{A}}

\newcommand{\fold}{\mathrm{fold}} 
\newcommand{\snap}{\mathrm{snap}} 
\newcommand{\start}{\mathrm{start}} 
\newcommand{\dL}{\Delta L}
\newcommand{\tbar}{\mathfrak{T}}
\newcommand{\Tsnap}{\tbar_\snap}
\newcommand{\Ai}{\mbox{Ai}}
\newcommand{\dMuEff}{\Delta \mu_{\mathrm{eff}}}
\newcommand{\pdd}[2]{\frac{\partial^{2} {#1}}{\partial {#2}^{2}}}
\newcommand{\pd}[2]{\frac{\partial #1}{\partial #2}}

\newcommand{\ddd}[2]{\frac{\upd^{2} {#1}}{\upd {#2}^{2}}}
\newcommand{\dd}[2]{\frac{\upd #1}{\upd #2}}

\journal{Journal of the Mechanics and Physics of Solids}

\begin{document}

\begin{frontmatter}

\title{Delayed bifurcation in elastic snap-through instabilities}


\author{Mingchao Liu$^{\dag}$, Michael Gomez$^{\ddag}$ and Dominic Vella$^{\dag}$}

\address{$^{\dag}$\:Mathematical Institute, University of Oxford, Woodstock Rd, Oxford, OX2 6GG, UK\\
$^{\ddag}$\:Department of Applied Mathematics and Theoretical Physics, University of Cambridge, Wilberforce Rd, Cambridge, CB3 0WA, UK}


\begin{abstract}
We study elastic snap-through induced by a control parameter that evolves dynamically. In particular, we study an elastic arch subject to an end-shortening that evolves linearly with time, i.e.~at a constant rate. For large end-shortening the arch is bistable but, below a critical end-shortening, the arch becomes monostable. We study when and how the arch transitions between states  and show that the end-shortening at which the fast `snap' happens depends on the rate at which the end-shortening is reduced. This lag in snap-through is a consequence of delayed bifurcation and occurs even in the perfectly elastic case when viscous (and viscoelastic) effects are negligible. We present the results of numerical simulations to determine the magnitude of this lag as the loading rate and the importance of external viscous damping vary. We also present an asymptotic analysis of the geometrically-nonlinear problem that reduces the salient dynamics to that of an ordinary differential equation; the form of this reduced equation is generic for snap-through instabilities in which the relevant control parameter is ramped linearly in time. Moreover, this asymptotic reduction allows us to derive analytical results for the observed lag in snap-through that are in good agreement with the numerical results of our simulations. Finally, we discuss scaling laws for the lag that should be expected in other examples of delayed bifurcation in elastic instabilities.
\end{abstract}




\end{frontmatter}

\section{Introduction}

The snap-through of an elastic object between two stable states is a striking demonstration of the rapid release of stored elastic energy. Indeed, so effective is snap-through that plants such as the Venus flytrap use it to rapidly release energy that is stored slowly through the growth and swelling of tissues, thereby catching their prey unawares \cite[][]{forterre2005,Skotheim2005}. In the same way, the hummingbird uses snap-through to generate fast motion in its beak and hence catch flying insects \cite[][]{Smith2011}. Taking inspiration from the natural world, a number of applications have sought to make use of snap-through instabilities \cite[][]{hu2015buckling}. For example, \cite{Holmes2007} demonstrated microscopic lenses whose curvature (and hence optical properties) could be rapidly changed via snap-through, while ventricular assist devices use snap-through as a temporary pump to aid the heart while a donor is sought \cite[][]{Goncalves2003}. By combining a series of snapping elements, a snapping metamaterial \cite[as introduced by][]{Rafsanjani2015} may find applications as diverse as the isolation and damping of vibrations or as a means of shape change in soft robots \cite[][]{Janbaz2020}.


In both technological and natural applications, it is the rapid motions associated with snap-through that are particularly useful; in particular, these motions can cause global shape changes while remaining in the elastic regime, and hence are highly reproducible. However, in many scenarios, simple estimates of the speed of snap-through lead to significant over-estimates: snap-through occurs much more slowly than would be expected by a naive balance between elastic and inertial forces \cite[][]{forterre2005}. Hence, this discrepancy is often attributed to a source of energy dissipation, such as viscous damping (either external or internal e.g.~poroelasticity) or viscoelasticy of the material. Indeed, several studies have shown that viscoelasticity may lead to a system that is actually monostable (when fully relaxed) temporarily appearing to be bistable, so that it undergoes a slow creeping motion before rapidly snapping between states --- what is referred to as temporary bistability \cite[][]{Santer2010}, pseudo-bistability \cite[][]{brinkmeyer2012,brinkmeyer2013,Gomez2019} or acquired bistability \cite[][]{Urbach2020}. 

While viscoelastic effects are undoubtedly important in numerous systems, the structure of the snap-through transition has also been shown to slow the dynamics significantly \cite[][]{pandey2014,gomez2017critical,Sano2018}, even in the absence of viscous effects. Generically, the transition corresponds to a saddle-node (fold) bifurcation in which the current equilibrium state abruptly ceases to exist as the relevant control parameter is varied, without first becoming unstable 
 \cite[this kind of bifurcation is also referred to as a limit-point instability in the engineering literature; see][for example]{bushnell1981}.  When the system is close to the bifurcation point, the behaviour is then heavily influenced by this closeness: since an equilibrium lies close by in parameter space, the dynamics is not as fast as might naively be expected. Broadly speaking, because the various forces within the structure are in balance at the bifurcation point (as this is an equilibrium solution), the net force on the structure is very small and its motions slow down significantly. This closeness to bifurcation \cite[called the `ghost' of the saddle-node bifurcation by][]{strogatz} causes a bottleneck for the dynamics of snap-through --- in general, the duration of this bottleneck dominates the overall snap-through time, and diverges as the bifurcation point is approached.

\begin{figure}[ht]
\centering
\vspace{0.5cm}
\includegraphics[width=0.9\textwidth]{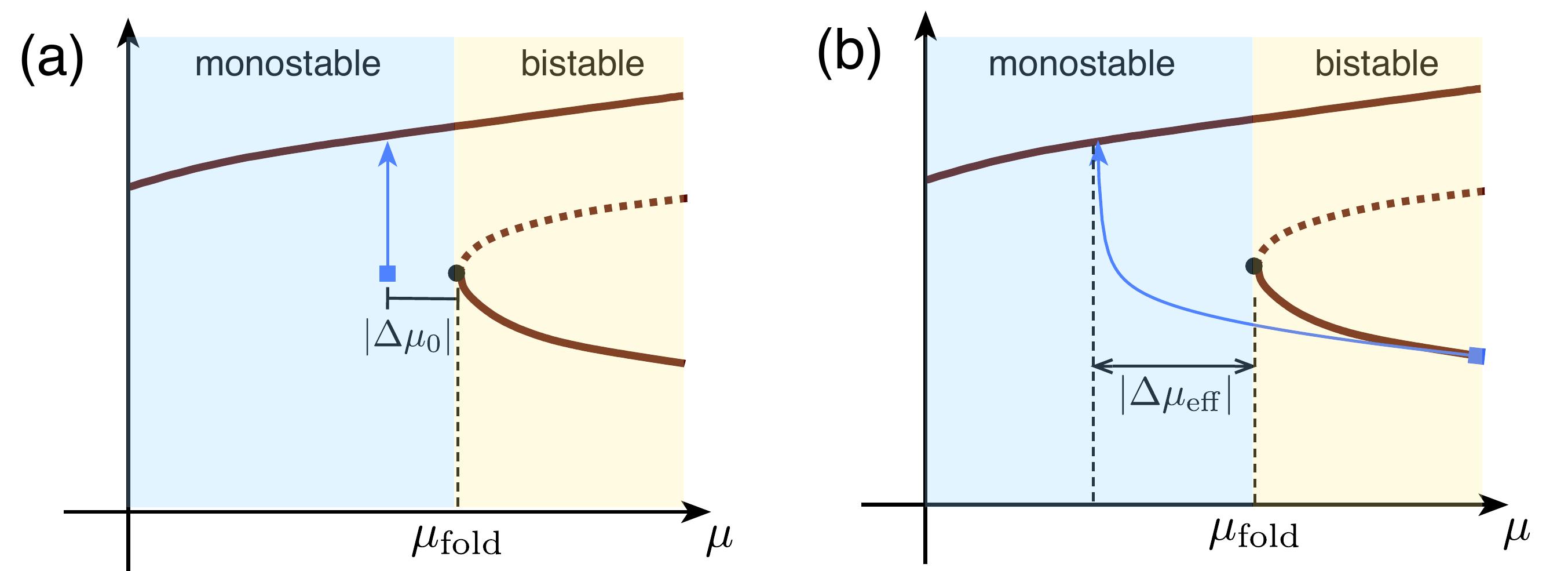}
\caption{Schematic showing the bifurcation structure that generically underlies elastic snap-through: for different values of a control parameter, $\mu$ (horizontal axis), an elastic structure is either bistable or monostable. (Here the vertical axis is a scalar `amplitude' that describes the equilibrium states, e.g.~the maximum displacement.) One of the stable states (solid curve) disappears at a saddle-node bifurcation, labelled $\mu=\mu_\fold$ (drawn as a filled circle), where it merges with an unstable state (dotted curve). In each panel the dynamics of snap-through depend on the path taken in parameter space, illustrated by the blue arrows, which begin at the filled square. (a) \cite{gomez2017critical} considered the dynamics with a fixed value of the bifurcation parameter $\mu=\mu_\fold+\Delta\mu_0$, with $\Delta\mu_0 < 0$, by starting from a shape close to the equilibrium shape at the bifurcation point. (b) In this paper, we consider an alternative scenario in which the system begins in a stable equilibrium state well before the bifurcation point. The parameter $\mu$ then evolves  in time at a finite rate. In particular, we seek to determine the delay in the control parameter at snap-through, labelled $|\dMuEff|$, observed with linear parameter variation, i.e.~$\mu=\mu_\fold+\mudot \:t$.}
\label{fig:BifurcationsSchematic}
\end{figure}

\cite{gomez2017critical} studied the effect of the proximity to bifurcation on the duration of snap-through in purely elastic systems. By starting the system a fixed distance $|\Delta\mu_0|$ beyond the bifurcation point $\mu_\fold$ in parameter space (see schematic in fig.~\ref{fig:BifurcationsSchematic}a), \cite{gomez2017critical} were able to determine experimentally and theoretically the duration of the bottleneck phase and hence the total snap-through time. In practice, these initial conditions were achieved using an external constraint that was suddenly removed to initiate snap-through. Extending this work, \cite{gomezthesis} considered the influence of external viscous damping on the snap-through dynamics. This revealed two possible regimes depending delicately on the importance of damping compared to the `distance' $|\Delta\mu_0|$, with different scaling laws for the snap-through time; in particular, the snap-through time was found to scale as $|\Delta\mu_0|^{-1/4}$ in the underdamped limit and $|\Delta\mu_0|^{-1/2}$ in the overdamped limit.

In many natural and man-made systems, however, the relevant bifurcation parameter that causes snap-through evolves smoothly  in time, so that the system evolves along a trajectory qualitatively similar to that shown in fig.~\ref{fig:BifurcationsSchematic}b.  For example, when stimulated by an insect, the leaf of the Venus flytrap gradually changes its intrinsic curvature in response to changes in turgor pressure; this change of curvature eventually makes the `open' configuration unstable and the leaf snaps `shut' \cite[][]{forterre2005}. Similarly, in many systems the stimulus is driven by mass transport and hence changes over a diffusion-limited timescale; for example, the snap-through of colloidal particles may be caused by a change in pH \cite[][]{epstein2015}, while hydrogel bilayers may snap upon solvent uptake \cite[][]{lee2010}. Moreover, in technological applications, the external loading that causes snap-through is also often controlled dynamically, for example with pneumatic pressure \cite[][]{Goncalves2003,Holmes2007}, magnetic forces \cite[][]{Loukaides2014,Seffen2016} and fluid loading \cite[][]{gomez2017b,Arena2017}. 

In these scenarios, because the effective bifurcation parameter is time varying, the analysis of \cite{gomezthesis} --- in which the system is quasi-statically placed beyond the snap-through transition --- is no longer applicable. Several questions then naturally arise: at which effective bifurcation parameter value, $\mu_\fold-|\dMuEff|$ (see fig.~\ref{fig:BifurcationsSchematic}b), does the snap-through take place? What determines the duration of snap-through in these evolving scenarios?

In low-dimensional systems, previous work has demonstrated how dynamic loading can delay the onset of instability and modify the scaling law describing bottleneck behaviour \cite[][]{tredicce2004,majumdar2013} --- a so-called delayed bifurcation \cite[][]{Su2001}. It is therefore of interest to understand how such effects influence the dynamics of snap-through. This may yield new insights into how systems such as the Venus flytrap behave near critical transitions and so is the main aim of the present paper.  One of the key challenges, however, is to extend the work on low-dimensional systems to an elastic continuum with infinitely many degrees of freedom, whose evolution is described by partial differential equations (PDEs).

In this paper we answer these questions through a careful study of a model system: the snap-through of an elastic arch that is subject to a time-dependent end-shortening (see fig.~\ref{fig:Setup}). For simplicity, we neglect viscoelasticity and consider the arch to be purely elastic (though we do consider external viscous damping with constant damping coefficient as a lumped model for energy dissipation via viscous effects). The remainder of this paper is organised as follows. In \S\ref{sec:TheoreticalFormulation} we present the geometrically-nonlinear (elastica) equations governing the dynamics of the arch and their non-dimensionalization. We also discuss a simplification of these equations, referred to as the quasi-linear approximation, which is used to obtain analytical results throughout the paper. We consider equilibrium solutions in \S\ref{sec:NonlinearStatics}, showing that the quasi-linear approximation captures well the bifurcation behaviour of the fully-nonlinear problem as the end-shortening is varied quasi-statically. In \S\ref{sec:NumericalDetails} we then present numerical results of the fully-nonlinear, dynamic problem, focussing on the case when the end-shortening varies linearly in time. We show that the system is subject to delayed bifurcation: the trajectory `lags' behind the quasi-static prediction in a way that depends subtly on the loading rate and a dimensionless measure of viscous damping. In \S\ref{sec:Asymptotics}, we seek to explain this behaviour in the framework of the quasi-linear approximation: we use an asymptotic reduction of the governing PDE to a single ordinary differential equation (ODE) that is valid near the snap-through threshold. We solve this ODE in \S\ref{sec:AmpEqnAnalysis}, before comparing its predictions to numerical results obtained for the fully-nonlinear, dynamic problem in \S\ref{sec:compareelastica}. Finally, in \S\ref{sec:SummaryDiscussion}, we summarize our findings and discuss other examples of delayed bifurcation in elastic instabilities.

\section{Theoretical formulation}
\label{sec:TheoreticalFormulation}

\subsection{Problem statement}

\begin{figure}[ht]
\centering
\vspace{0.5cm}
\includegraphics[width=0.65\textwidth]{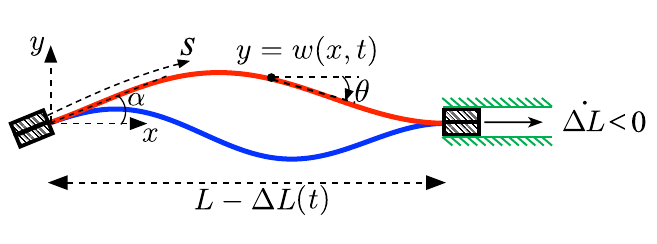}
\caption{A schematic diagram of the problem considered in this paper. An elastic strip of length $L$ is confined by two clamps that are separated by a horizontal distance $L-\dL$ (the strip's width is directed into the page) forming an arch. One clamp imposes $\theta = \alpha$, while the other imposes $\theta=0$; here $\theta$ is the inclination of the strip to the horizontal. For a given value of $\alpha$, the arch may be bistable or monostable depending on the value of $\dL$: in the bistable scenario it can adopt either the natural state (red curve) or the inverted state (blue curve). If monostable, it can only adopt the natural state. In this paper we consider the dynamics of the snap-through from the inverted to natural states resulting from variation in $\dL$ at a constant rate $\dot{\dL}<0$ (as indicated by the arrow).}
\label{fig:Setup}
\end{figure}

We consider a linearly elastic strip of density $\rho_s$, Young's modulus $E$, thickness $h$ and length $L$, with $h \ll L$. The ends of the strip are a distance $L-\dL$ apart, so that there is an end-end displacement $\dL>0$, which causes the strip to buckle out of plane forming an arch. To obtain a system that is bistable for some parameter values, but monostable for others, we follow \cite{gomez2017critical} in the choice of conditions that are imposed at the ends of the strip: one end of the strip is inclined at a constant angle $\alpha$ to the horizontal, while the other end is clamped horizontally. A sketch of the setup is shown in fig.~\ref{fig:Setup}. In the static case, it is known \cite[][]{gomez2017critical} that for sufficiently small $\alpha$ (or large $\dL$) the arch is bistable while for sufficiently large $\alpha$ (or small $\dL$) the arch is monostable. Since we are interested in the dynamics of the snap-through transition, we consider how the arch evolves from the stable equilibrium that disappears at the transition from bistability to monostability. In particular, we imagine the angle $\alpha$ to be fixed and allow the end-shortening $\dL$ to vary, so that in the static case the arch is monostable or bistable if $\dL<\dL_c$ or $\dL>\dL_c$, respectively, where $\dL_c$ is a critical value that depends on the angle of inclination, $\alpha$. In this paper, we  shall focus on the effect of a dynamically evolving confinement, i.e.~$\dL=\dL(t)$.

\subsection{Governing equations}

The deformed centreline of the arch can be expressed as
\begin{equation}
\boldsymbol{r} (s,t)=x(s,t)\:\boldsymbol{e}_x+y(s,t)\:\boldsymbol{e}_y,
\end{equation}
\noindent
where $\boldsymbol{e}_x$ and $\boldsymbol{e}_y$ are unit vectors in the $x$ and $y$ directions, respectively (see fig.~\ref{fig:Setup}), $s$ is the arc length measured along the centreline from the left clamp, and $t$ is time. We assume that the strip undergoes inextensible, unshearable deformations (due to its slenderness $h \ll L$) in the $x$-$y$ plane, i.e.~perpendicular to its width. In this intrinsic coordinate system, the tangent vector is then given by 
\begin{equation}
\frac{\partial\boldsymbol{r}}{\partial s} = \cos\theta \:\boldsymbol{e}_x + \sin\theta \:\boldsymbol{e}_y.
\end{equation} 
We write $\boldsymbol{n}(s,t)$ for the resultant force and $\boldsymbol{m}(s,t)$ for the resultant torque (per unit width) attached to the centreline, obtained by averaging the internal force/torque over the cross-section. Force balance on the strip, subject to a viscous damping per unit area of strip, $\Upsilon$ (assumed constant for simplicity), gives
\begin{equation}
\frac{\partial \boldsymbol{n}}{\partial s} = \rho_s h \frac{\partial^2 \boldsymbol{r}}{\partial t^2} + \Upsilon \frac{\partial \boldsymbol{r}}{\partial t},
\label{eqn:ForceBal}
\end{equation} 
while torque balance on the strip gives
\begin{equation}
\frac{\partial \boldsymbol{m}}{\partial s} + \frac{\partial \boldsymbol{r}}{\partial s} \times \boldsymbol{n} = \rho_s I \frac{\partial^2 \theta}{\partial t^2} \boldsymbol{e}_z.
\label{eqn:TorqueBal}
\end{equation}
Here $I=h^3/12$ is the moment of inertia of the strip (per unit width). We also make the constitutive assumption (the Euler-Bernoulli constitutive law) that the restoring torque caused by the strip's bending stiffness, $B=EI$, is
\begin{equation}
\boldsymbol{m} = B \frac{\partial \theta}{\partial s} \boldsymbol{e}_z.
\label{eqn:MomentConst}
\end{equation} 
Substituting \eqref{eqn:MomentConst} into the torque balance, \eqref{eqn:TorqueBal}, and writing $\boldsymbol{n} = n_x \boldsymbol{e}_x + n_x \boldsymbol{e}_y$, gives the dynamic elastica equation
\beq
B\frac{\partial^2\theta}{\partial s^2} -n_x \sin\theta + n_y \cos\theta =\rho_sI\frac{\partial^2\theta}{\partial t^2},
\eeq 
which is a second order equation for $\theta(s,t)$ once the force components are determined from \eqref{eqn:ForceBal}.  As well as suitable initial conditions, we require boundary conditions, which we discuss now.

\subsection{Boundary conditions}

The boundary conditions imposed at the clamped ends are
\begin{equation}
\theta(0,t)=\alpha,\ \theta(L,t)=0.
\label{eqn:ThetaBCdimension}
\end{equation}
We take the origin of our two-dimensional coordinate system to be at the inclined clamp (see fig.~\ref{fig:Setup}), i.e.~we let
\begin{equation}
x(0,t)=0,\ y(0,t)=0,
\label{eqn:XYBC1dimension}
\end{equation} 
and assume that the second (horizontal) clamp is at the same vertical level, i.e.
\beq
y(L,t)=0.
\eeq
The final boundary condition comes from the applied end-shortening, $\Delta L(t)$, which enters via the evolving $x$-position of the horizontal clamp:
\begin{equation}
x(L,t)=L-\Delta L(t).
\label{eqn:XYBC2dimension}
\end{equation}
For the moment, $\dL(t)$ will be allowed to be a general function of time, $t$. Ultimately, however, we shall consider a constant rate of change of $\dL$, i.e.~$\dL(t) = \dL(0)+\dot{\dL}\: t$.

\subsection{Non-dimensionalization}

For simplicity, we scale all lengths by the total length of the strip, $L$, all  forces within the strip  by the characteristic buckling load $N^\ast=B/L^2$ and time by the inertial timescale $t^\ast=(\rho_shL^4/B)^{1/2}$. We use upper case letters to denote dimensionless variables (where this does not conflict with standard notation), i.e.~we write
$$
(\boldsymbol{r},s,x,y)=L (\boldsymbol{R},S,X,Y), \quad t = t^\ast T, \quad (n_x,n_y)=\frac{B}{L^2}(N_X,N_Y).
$$
The full system of dimensionless equations becomes
\begin{equation}
\frac{\partial X}{\partial S} = \cos \theta,\ \frac{\partial Y}{\partial S} = \sin \theta,
\label{eqn:GeomeRL}
\end{equation}

\begin{equation}
\frac{\partial N_X}{\partial S} = \frac{\partial^2 X}{\partial T^2} + \nu \frac{\partial X}{\partial T},
\label{eqn:HFBalND}
\end{equation}

\begin{equation}
\frac{\partial N_Y}{\partial S} = \frac{\partial^2 Y}{\partial T^2} + \nu \frac{\partial Y}{\partial T},
\label{eqn:VFBalND}
\end{equation}

\begin{equation}
\frac{\partial^2 \theta}{\partial S^2} - N_X \sin\theta + N_Y \cos\theta = \mathcal{S} \frac{\partial^2 \theta}{\partial T^2},
\label{eqn:TorqueBalND}
\end{equation}
\noindent
where our non-dimensionalization introduces two dimensionless parameters:
\beq
\nu = \frac{\Upsilon L^2}{\sqrt{\rho_s h B}}, \quad \mathcal{S} = \frac{h^2}{12L^2}.
\label{eqn:NonDimPars}
\eeq 
The first of these dimensionless parameters, $\nu$, is a dimensionless damping coefficient, which measures the ratio of viscous forces per unit area ($\sim \Upsilon L/t^\ast$) to bending forces ($\sim B/L^3$) over the inertial timescale $t^\ast$, and will be important in determining the dynamics of the system. The second dimensionless parameter, ${\cal S}$, is a purely geometrical parameter that measures the slenderness of the strip. This slenderness parameter is of the order of $10^{-8}$ for a piece of paper, while in the experiments of \cite{gomez2017critical}, ${\cal S} \in (5.5, 18)\times 10^{-8}$ (for PTE strips) and  ${\cal S} \in (1.1, 4.3)\times 10^{-8}$ (for steel strips). Since in these examples, ${\cal S}\ll 1$, we neglect the corresponding term in \eqref{eqn:TorqueBalND}, which represents the rotational inertia of an element of the strip: individual elements are in quasi-static torque balance, even as the components of the force evolve\footnote{We note that any temporal `boundary' layers or transients, in which rotational inertia is important, are expected to have duration $T = O({\cal S}^{1/2})$ based on a balance between terms in \eqref{eqn:TorqueBalND}. Because of the extremely small values of ${\cal S}$ typically encountered, these transients are much faster than the timescale that we consider for snap-through, which is $O(1)$ or larger, depending on the importance of damping, since we have non-dimensionalized time by the inertial timescale. Moreover, as discussed further in \S\ref{sec:NumericalDetails}, we only consider relatively slow variations in the end-shortening. Hence we neglect these transients and suppose that the strip is always in quasi-static torque balance.}. With this simplification, \eqref{eqn:TorqueBalND} becomes
\beq
\frac{\partial^2 \theta}{\partial S^2} = N_X \sin\theta - N_Y \cos\theta.
\label{eqn:ElasticaND}
\eeq
The boundary conditions in terms of dimensionless variables become
\begin{equation}
\theta(0,T)=\alpha,\ \theta(1,T)=0,
\label{eqn:ThetaBC}
\end{equation} 
together with
\begin{equation}
X(0,T)=Y(0,T)=Y(1,T)=0,
\label{eqn:XYBC1}
\end{equation}
and
\begin{equation}
X(1,T)=1-d(T),
\label{eqn:XYBC2}
\end{equation} 
where the (time-dependent) end-shortening imposed on the arch is
\begin{equation}
d(T)=\frac{\Delta L(T)}{L}.
\label{eqn:dDefn}
\end{equation} 
 Integrating the geometric relations \eqref{eqn:GeomeRL}, the above boundary conditions at $S = 1$ enforcing the position of the horizontal clamp can be rephrased as
\beq
\int_0^1\cos\theta~\upd S =1-d(T), \quad \int_0^1\sin\theta~\upd S=0.
\label{eqn:IntConstraints}
\eeq 
Finally, the system of equations is closed by appropriate initial conditions, which we discuss later in \S\ref{sec:NumericalDetails}.

\subsection{Quasi-linear approximation}

In this paper we shall study the numerical solutions of the system of equations \eqref{eqn:GeomeRL}--\eqref{eqn:VFBalND} and \eqref{eqn:ElasticaND}, with boundary conditions \eqref{eqn:ThetaBC} and integral constraints \eqref{eqn:IntConstraints}. However, to interpret these numerical solutions, we require a simplified system for which it is possible to make some analytical progress. Linearization is a common means of doing this but a linear system cannot, in general,  have a controlled amplitude and hence cannot exhibit snap-through. Instead, we linearize the problem but retain the leading-order nonlinearity in the constraints \eqref{eqn:IntConstraints}.

We begin by linearizing the elastica equation \eqref{eqn:ElasticaND} by assuming that all angles are small, i.e. $\theta\ll1$. In this limit the angle $\theta$ is approximately equal to the local slope of the arch, i.e.~$\theta\approx\partial Y/\partial X\ll1$. We may then approximate $\cos\theta\approx1$, $\sin\theta\approx\tan\theta=\partial Y/\partial X$ etc.~so that the moment balance \eqref{eqn:ElasticaND} becomes
\beq
\frac{\partial^3 Y}{\partial X^3} -N_X\frac{\partial Y}{\partial X}+N_Y=0.
\label{eqn:ElasticaLin}
\eeq 
Using \eqref{eqn:GeomeRL}, we have $S\approx X$ and so \eqref{eqn:HFBalND} shows that $N_X$ depends only on $T$, not $S$ ($\approx X$). The vertical force $N_Y$ can be eliminated by differentiating \eqref{eqn:ElasticaLin} with respect to $X$ and using \eqref{eqn:VFBalND}. We therefore recover the usual dynamic beam equation:
\beq
\frac{\partial^2Y}{\partial T^2}+\nu \frac{\partial Y}{\partial T} + \frac{\partial^4 Y}{\partial X^4} -N_X\frac{\partial^2 Y}{\partial X^2}=0.
\label{eqn:ElasticaLinDyn}
\eeq The boundary conditions \eqref{eqn:ThetaBC}--\eqref{eqn:XYBC1} become
\beq
Y(0,T)= 0, \quad \left.\frac{\partial Y}{\partial X}\right|_{X=0}=\alpha, \quad Y(1,T)=\left.\frac{\partial Y}{\partial X}\right|_{X=1}=0.
\label{eqn:LinBCs}
\eeq
 Equations \eqref{eqn:ElasticaLinDyn}--\eqref{eqn:LinBCs} represent a fifth-order system (since the horizontal force $N_X$ is not known \emph{a priori}) with four boundary conditions. To close the system, we retain the first non-trivial term in the expansion of the end-shortening constraint  \eqref{eqn:IntConstraints} for $\theta \ll 1$, which gives
\beq
d(T)=1-\int_0^1\cos\theta~\upd S\approx \frac{1}{2}\int_0^1\left(\frac{\partial Y}{\partial X}\right)^2~\upd X.
\label{eqn:LinConstraint}
\eeq

We shall refer to the system \eqref{eqn:ElasticaLinDyn}--\eqref{eqn:LinConstraint} as the quasi-linear dynamic problem. We shall see that it is possible to make significant progress in understanding the behaviour of this system close to the snap-through threshold. For now, one important illustration of the insight afforded by the study of the quasi-linear system can be seen simply by noting that, upon rescaling $Y$ with the imposed clamp angle, i.e.~letting
\beq
Y(X,T)=\alpha W(X,T),
\label{eqn:Wscaling}
\eeq 
the problem depends only on the dimensionless damping $\nu$ and a new parameter (which enters via \eqref{eqn:LinConstraint})
\beq
\mu(T)=\frac{d(T)}{\alpha^2}.
\label{eqn:MuDefn}
\eeq  
The parameter $\mu$ measures the typical arch angle induced by the imposed compression (which would be expected to scale with $d^{1/2}=(\dL/L)^{1/2}$) in comparison to the angle imposed by the boundary conditions, $\alpha$. We note also that our definition of $\mu$ is equivalent to that used by  \cite{gomez2017critical}, which in our notation is given by $\alpha/d^{1/2}$; we use a different form simply because in the dynamic case we wish to vary the end-shortening $d(T)$, and so we choose a form of $\mu$ proportional to $d$ that makes this more natural.

Having derived the dimensionless equations of motion, and their quasi-linear approximation, in the next section we demonstrate the importance of the parameter $\mu$ by considering the static behaviour of the arch.

\section{Static bifurcation behaviour}
\label{sec:NonlinearStatics}

In the absence of any time-dependence, the force balances \eqref{eqn:HFBalND} and \eqref{eqn:VFBalND} imply that the horizontal and vertical force components within the arch, $N_X$ and $N_Y$, are constant; these constants are determined by solving the moment balance \eqref{eqn:ElasticaND} together with the boundary conditions \eqref{eqn:ThetaBC} and the global constraints \eqref{eqn:IntConstraints}, where now $d$ is independent of $T$.

Note that in this fully-nonlinear problem, the behaviour of the arch depends on the values of $\alpha$ and $d=\Delta L/L$ independently, and not simply on the parameter $\mu$ (defined in \eqref{eqn:MuDefn}) as with the quasi-linear problem. In the numerical results presented here, we fix a positive value of the clamp angle, $\alpha>0$, and vary the dimensionless end-shortening $d$ quasi-statically. For each value of $\alpha$ and $d$, we solve the resulting boundary-value problem (i.e.~\eqref{eqn:ElasticaND} subject to \eqref{eqn:ThetaBC} and \eqref{eqn:IntConstraints}) numerically by discretizing in $S$ with $N=100$ grid points and solving the resulting set of algebraic equations\footnote{We use the same spatial discretization later when solving the dynamic equations, so that the static solutions described here are exact equilibrium solutions of the discretized dynamic equations; for details of the discretization scheme see \S\ref{sec:NumericalDetails} below and also \ref{sec:AppendixNums}.}. Once the discretized solution for $\theta(S)$ and the force components are determined, the geometric relations in \eqref{eqn:GeomeRL} can be integrated to determine the equilibrium shape of the arch, $\boldsymbol{R}(S) = X(S)\boldsymbol{e}_X + Y(S)\boldsymbol{e}_Y$. 

\begin{figure}
\centering
\vspace{0.2cm}
\includegraphics[width=0.9\textwidth]{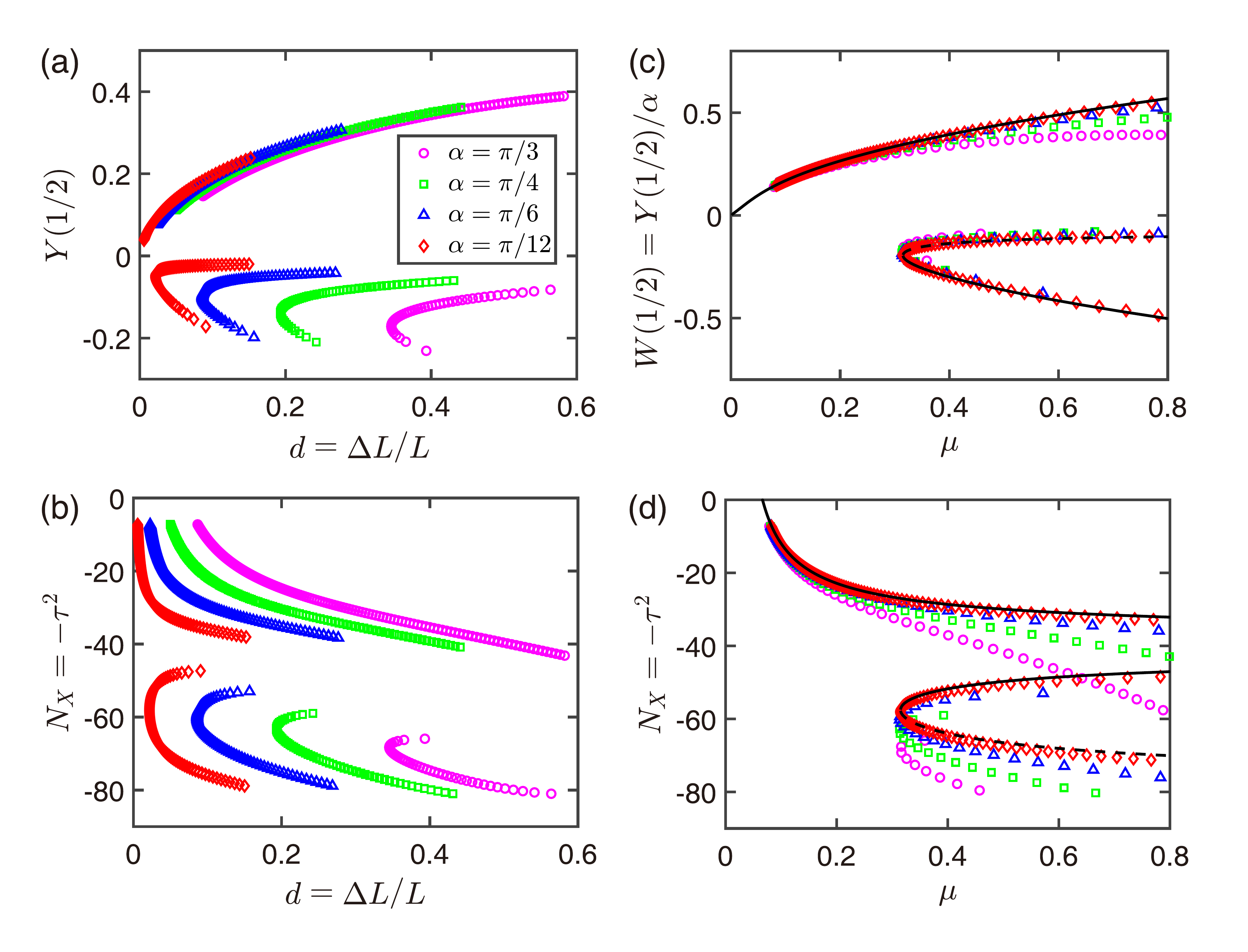}
\caption{The steady bifurcation behaviour of the arch (determined from solutions of the moment balance \eqref{eqn:ElasticaND} subject to \eqref{eqn:ThetaBC} and \eqref{eqn:IntConstraints}). Presented are raw numerical results for (a) the vertical displacement of the midpoint, $Y(1/2)$, and (b) the horizontal force in the arch, $N_X$, for different angles of inclination, $\alpha$ (as described in the legend). When rescaled as suggested by the quasi-linear theory, these raw data collapse onto a universal bifurcation structure as $\alpha\to0$. The predictions of the quasi-linear model (described by eqns \eqref{eqn:LinMidPoint} and \eqref{eqn:LinConstraintDetail}) are also shown as black curves (the dashed branch is unstable). (Note that in (c) the unstable branch lies above the fold point, but in (d) the unstable branch lies below the fold point and has the largest compressive force $-N_X=\tau^2$.) }
\label{fig:StaticBifn}
\end{figure}

To track the bifurcation behaviour of the system, we plot the vertical position of the arch midpoint, $Y(1/2)$, as a function of the relative end-shortening, $d$; this plot is  shown for four values of $\alpha$ in fig.~\ref{fig:StaticBifn}a. In each case, the corresponding values of the horizontal force component, $N_X$, are plotted in fig.~\ref{fig:StaticBifn}b. As anticipated, we find that for a given value of the angle $\alpha$, the system is bistable for $d>d_c(\alpha)$. As discussed further below, the upper branches in fig.~\ref{fig:StaticBifn}a, with positive midpoint displacement, are linearly stable and correspond to the natural equilibrium shape (drawn as a red curve in fig.~\ref{fig:Setup}); the branch in fig.~\ref{fig:StaticBifn}a with negative midpoint displacement and lying below the fold point is also stable and corresponds to the inverted equilibrium shape (blue curve in fig.~\ref{fig:Setup}). The third branch, connected above the fold point in fig.~\ref{fig:StaticBifn}a, is linearly unstable and never observed in our dynamic simulations. (We note that the location of the stable inverted and unstable branches are inverted in fig.~\ref{fig:StaticBifn}b, so that the inverted shape corresponds to the branches above the fold point there.) If $d<d_c$, only the stable natural shape then exists: the system is monostable. 

To understand the behaviour observed in these bifurcation diagrams a little better, we now consider the static solutions of the quasi-linear problem.

\subsection{Static behaviour of the quasi-linear problem \label{sec:QuasiLinStatics}}

Recalling the rescaled variable $W(X)=Y(X)/\alpha$, the static solution of the  beam equation, \eqref{eqn:ElasticaLinDyn}, subject to the boundary conditions \eqref{eqn:LinBCs} may readily be shown to be
\begin{equation}
W(X)=\frac{\tau X(\cos\tau-1)+\tau [\cos\tau (1-X)-\cos\tau]-\sin\tau X-\sin\tau(1-X)+\sin\tau}{\tau(2\cos\tau+\tau \sin\tau-2)},
\label{eqn:EqmShapeLin}
\end{equation} 
where $N_X=-\tau^2$ is introduced for mathematical convenience. The constant $\tau$ is not, however, determined thus far; it must be chosen to ensure that the shape \eqref{eqn:EqmShapeLin} satisfies the imposed end-shortening, i.e.~\eqref{eqn:LinConstraint}. Substituting \eqref{eqn:EqmShapeLin} into \eqref{eqn:LinConstraint}, we find that
\begin{equation}
2\mu=\int^1_0\left(\frac{\upd W}{\upd X}\right)^2~\upd X=\frac{2{\tau}^3-{\tau}^2(\sin2\tau+4\sin\tau)+4\tau(\cos\tau-\cos2\tau)+2(\sin2\tau-2\sin\tau)}{4\tau(2\cos\tau+\tau \sin\tau-2)^2},
\label{eqn:LinConstraintDetail}
\end{equation}
where $\mu=d/\alpha^2$, as defined in \eqref{eqn:MuDefn}.

We emphasize that the quasi-linear approximation relies on the angle $\alpha\ll1$ (to ensure small slopes, $\theta\ll1$) but that, with this restriction, the parameter $\mu$ encodes the crucial geometrical features of the problem: the angle $\alpha$ and relative end-shortening $d$ do not enter the quasi-linear problem independently. To test the relevance of the parameter $\mu$ to the fully-nonlinear problem, each of fig.~\ref{fig:StaticBifn}c,d shows the numerical data of fig.~\ref{fig:StaticBifn}a,b re-plotted with the horizontal axis now the corresponding value of $\mu$. As might be expected, the numerical results accounting for finite clamp angles $\alpha$ collapse onto universal curves as $\alpha\to0$. 
 
To confirm that the universal curves observed in the numerical solutions as $\alpha\to0$ correspond to the quasi-linear prediction, for a given value of $\mu$ we solve \eqref{eqn:LinConstraintDetail} (which represents an equation for the dimensionless horizontal force $N_X=-\tau^2$ that must be applied to the arch to achieve a given relative end-shortening, $d$).  In practice, the value of $\tau$ for a given $\mu$ must be determined numerically using a root-finding algorithm, for example the \textsc{MATLAB} routine \texttt{fzero}.  To compute the bifurcation diagram, however, it is convenient to use $\tau$ as a control parameter and so we note that \eqref{eqn:EqmShapeLin} gives the vertical position of the arch's midpoint as
\beq
W(1/2)=\frac{Y(1/2)}{\alpha}=\frac{\tan(\tau/4)}{2\tau}.
\label{eqn:LinMidPoint}
\eeq 
This relation, together with \eqref{eqn:LinConstraintDetail}, allows us to superimpose the analytical prediction of the quasi-linear theory onto the numerical data in each of fig.~\ref{fig:StaticBifn}c,d (curves). These curves reproduce the numerical data in the limit $\alpha\to0$, with excellent agreement even for moderately small clamp angles $\alpha \lesssim \pi/6$. 

An important feature of the bifurcation diagrams computed from the quasi-linear theory is that, in general, there are multiple roots. These correspond to increasingly large compressive values of the force, $-N_X=\tau^2$. However, since larger values of $\tau$ correspond to more convoluted arch shapes with larger wave number (via the trigonometric functions in \eqref{eqn:EqmShapeLin}) that are likely to be unstable, we will be most interested in smaller values of $\tau$ in general. In fact, using the method developed by \cite{maddocks1987}, it is possible to infer stability of branches directly from the bifurcation diagram, once it is re-plotted in terms of `preferred' coordinates that can be determined by rephrasing the equilibrium equations as a variational problem. In this way, we find that it is the two branches with the smallest values of $\tau$ that are linearly stable, while the branches with larger values of $\tau$ are all unstable --- the two stable branches are shown as solid curves in each of fig.~\ref{fig:StaticBifn}c,d while the first unstable branch is shown as the dashed curve in fig.~\ref{fig:StaticBifn}c,d. 

As expected, the stable branch corresponding to the inverted shape disappears in a collision with the unstable branch at a critical value $\mu=\mu_\fold$: there is a saddle-node bifurcation \cite[see][for example]{strogatz}. The analytical expressions from the quasi-linear problem  allow us to determine the properties of the  saddle-node (fold) bifurcation to be:
\beq
\mu_{\fold} \approx 0.3150,\quad W_{\fold}(1/2) \approx -0.1951,\quad \tau_{\fold} \approx 7.5864.
\label{eqn:FoldProperties}
\eeq 
These values agree well with the position of the fold in the fully-nonlinear bifurcation diagrams (fig.~\ref{fig:StaticBifn}b,d). This success of the quasi-linear approach motivates us to use a similar approach to study the dynamic problem. First, however, we present  numerical results of the fully-nonlinear dynamic problem.

\section{Numerical solution of the dynamic problem}
\label{sec:NumericalDetails}

\subsection{Details of the numerical scheme}

To solve the fully-nonlinear dynamic problem \eqref{eqn:GeomeRL}--\eqref{eqn:VFBalND} and \eqref{eqn:ElasticaND} subject to the boundary conditions \eqref{eqn:ThetaBC} and integral constraints \eqref{eqn:IntConstraints}, we use the method of lines \cite[see][for example]{Schiesser2009}. More specifically, we discretize the arc length $S$ on a uniform mesh composed of $(N + 1)$ grid points in the interval $[0, 1]$. We use second-order centered differences to approximate the spatial derivatives appearing in \eqref{eqn:ElasticaND}, and we apply the trapezium rule to compute the integrals appearing in \eqref{eqn:IntConstraints}. In this way, the problem reduces to a system of differential-algebraic equations (DAEs), consisting of $(N+1)$ ODEs in time and algebraic constraints
that enforce \eqref{eqn:IntConstraints}. This system of DAEs can be solved in \textsc{MATLAB} using the routine \texttt{ODE15s}. We are able to demonstrate second-order accuracy in the convergence of our numerical scheme as $N$ is increased; further details of the scheme and convergence plots are provided in Appendix A. In all results presented below we take $N = 100$.

Throughout the remainder of this paper, we consider a constant rate of decrease in the relative end-shortening, i.e.~we write
\beq
d(T)=d_c(\alpha)+\dot{d}\: T,
\label{eqn:dLramp}
\eeq 
with $\dot{d}<0$; here we choose $T = 0$ when $d$ is equal the critical end-shortening at the saddle-node bifurcation, $d_c(\alpha)$, which depends on the angle of inclination $\alpha$.  To enable a later comparison with the quasi-linear theory, we focus on relatively slow ramping rates compared to the inertial timescale, i.e.~$|\dot{d}| \ll 1$. This also allows us to explore how the dynamics are affected by the snap-through bifurcation --- for larger values of $|\dot{d}|$, the dynamics simply become limited by arch inertia and do not display any delay behaviour.

While \eqref{eqn:dLramp} is implemented numerically, we shall present our results in terms of the quantities presented in the quasi-linear theory using, in particular, the parameter $\mu$ defined in \eqref{eqn:MuDefn}. The linear decrease in $d$ implies that
\beq
\mu(T)=\mu_{\fold}(\alpha)-|\mudot|T,
\eeq 
where $\mu_{\fold}(\alpha)$ denotes the bifurcation value of $\mu$ as determined from the fully-nonlinear static problem\footnote{Generally we drop the explicit $\alpha$-dependence for the value predicted by the quasi-linear theory, provided earlier in \eqref{eqn:FoldProperties}, so that $\mu_\fold \approx 0.3150 =\lim_{\alpha\to0}\mu_\fold(\alpha)$.} for a particular value of $\alpha$ and $|\mudot|=-\dot{d}/\alpha^2$.  We will plot results in terms of the distance beyond the fold point, which, using $\mu_\fold(\alpha)-\mu=|\mudot|T$, simply corresponds to a rescaled time.

As the initial condition, we impose that the arch starts in an inverted equilibrium state with a value of $\mu > \mu_\fold$ corresponding to a large negative time, $T_{\start}<0$. (In practice, we achieve this by imposing an end-shortening $d_{\start}$ that is significantly greater than the critical value $d_c(\alpha)$; this ensures that the arch begins well below the snap-through transition.) The corresponding inverted shape, $Y_{\start}(X)$, that is far away from the equilibrium shape at the fold, $Y_{\fold}(X)$, can be calculated by solving the static system of equations as described in \S\ref{sec:NonlinearStatics}; the corresponding  start time $T_{\start}<0$ can be readily inferred from \eqref{eqn:dLramp}.

\subsection{Numerical results}

\begin{figure}
\centering
\includegraphics[width=0.9\textwidth]{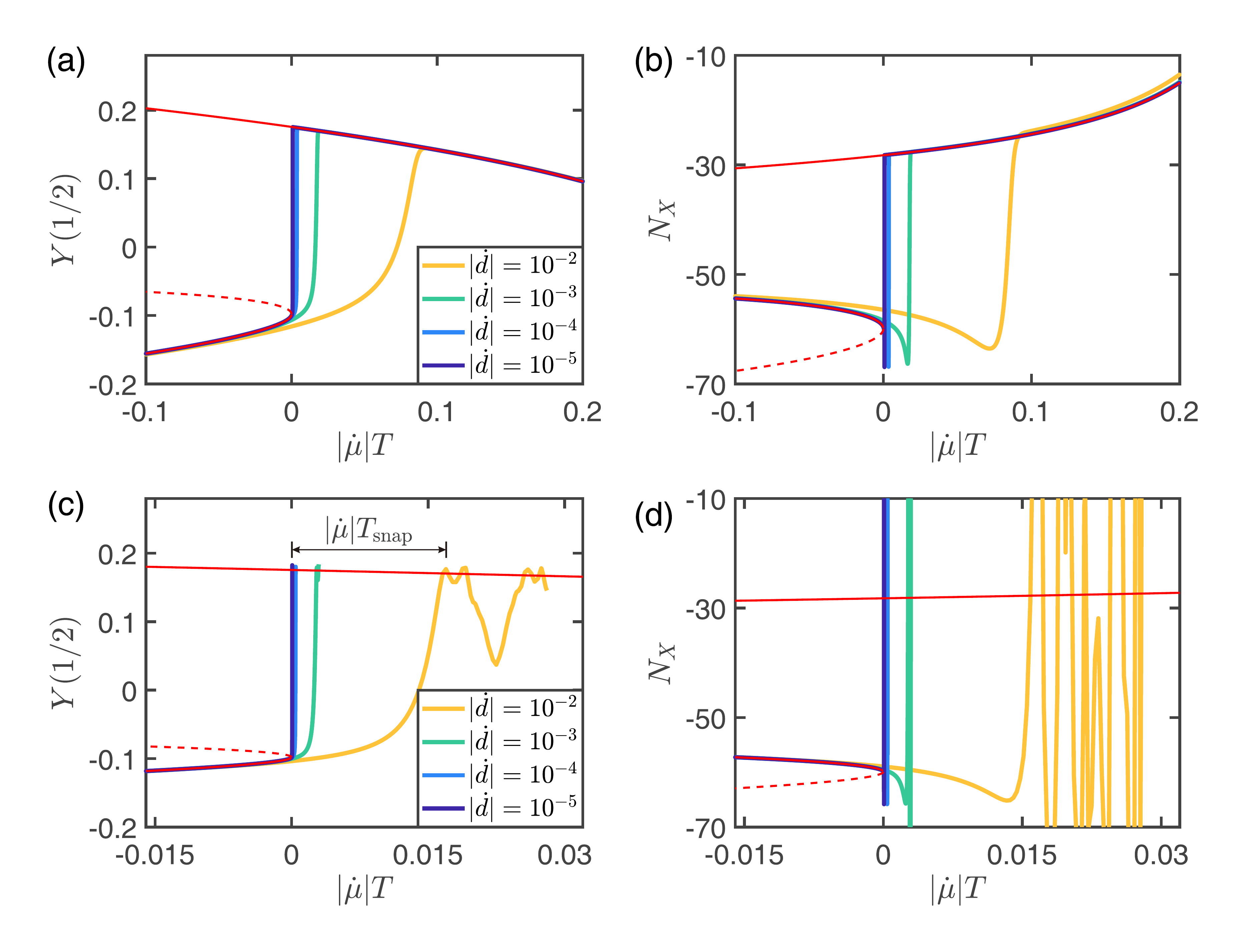}
\caption{The evolution of the midpoint displacement, $Y(1/2,T)$, and the horizontal force resultant, $N_X$, as functions of the distance beyond the quasi-statics snap-through threshold, $\mu_\fold(\alpha)-\mu(T)=|\mudot|T$. We vary the loading rates $\dot{d}=\dot{\dL} \: t_\ast/L=\alpha^2\dot{\mu}$ (shown by the legend) while fixing $\alpha=\pi/6$. In each plot, the results of dynamic simulations are shown for different loading rates $\dot{d}$ (given by the legend), together with the quasi-static bifurcation diagram (solid and dashed red curves). (a)--(b) Numerical results with large damping,  $\nu=100$, show that there is a delay in the transition between the stable solution branches as $\mu$ varies; the size of this delay depends on the loading rate. (c)--(d) Numerical results with a significantly smaller damping, $\nu=0.01$, show a qualitatively similar delay in the transition. (In this case snap-through is followed by significant underdamped oscillations; for clarity only the start of these are plotted.) The prevalence of this delay with both small and large damping shows that the phenomenology is a feature of the snap-through transition, rather than the presence of damping alone.
\label{fig:RawTrajectories}
}
\end{figure}

The results of numerical simulations with a moderately shallow clamp angle, $\alpha=\pi/6$, and a range of ramping rates, $|\dot{d}| \leq 10^{-2}$, are shown in fig.~\ref{fig:RawTrajectories}.  In particular, figs.~\ref{fig:RawTrajectories}a--b display results for overdamped dynamics with a large damping coefficient, $\nu = 100$, while figs.~\ref{fig:RawTrajectories}c--d display results for underdamped dynamics with $\nu = 0.01$. These results show that snap-through does not occur at the same effective value of $\mu$ independently of the rate of loading, $\dot{d}$: a significant delay in snap-through is observed, though the size of this delay decreases as $\dot{d}\to0$. Notably, a qualitatively similar delay is observed in both overdamped and underdamped scenarios.

\begin{figure}
\centering
\vspace{0.2cm}
\includegraphics[width=0.9\textwidth]{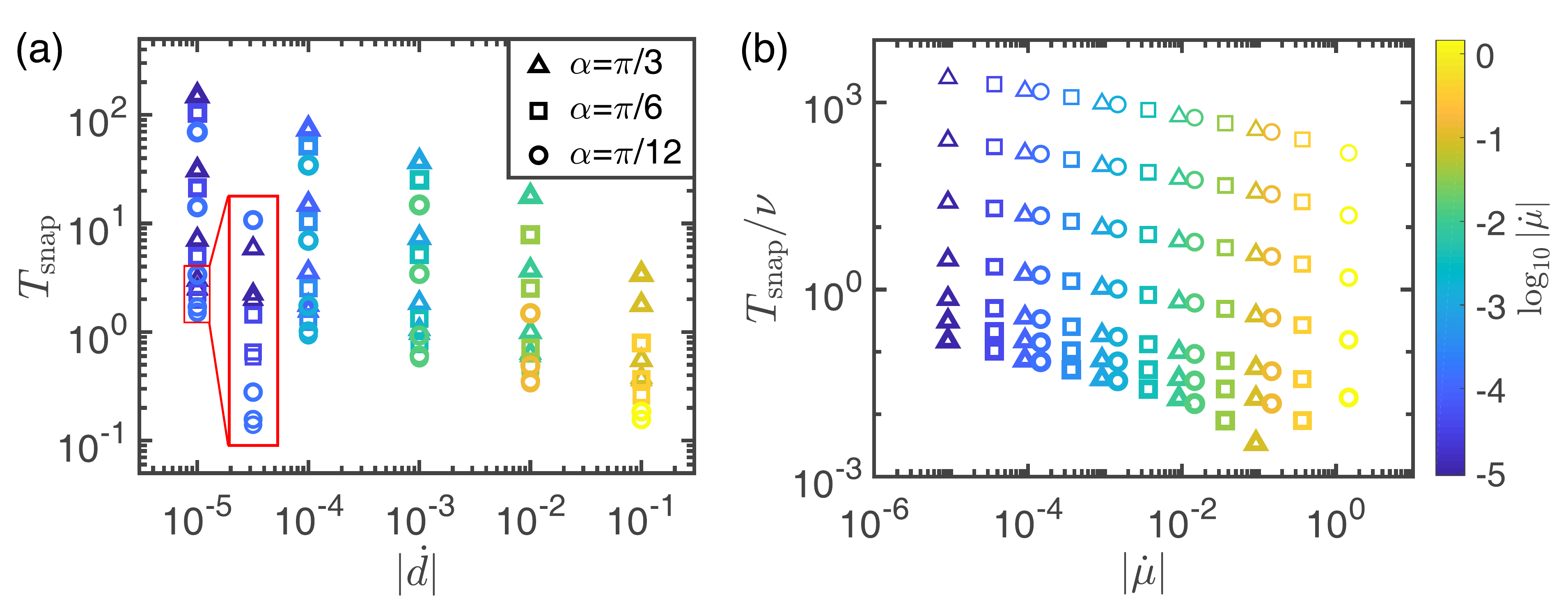}
\caption{The size of the delay in snap-through for different loading rates, $\dot{d}$, and damping coefficient, $\nu$ (indicated by symbol thickness, with the thinnest having $\nu=10^{-3}$ and each increment representing an increase by a factor of $10$ up to $\nu=10^3$ for the thickest symbols). Numerical results of the fully-nonlinear problem with different inclination angles, $\alpha$ (indicated by the symbol shape), are shown. Here the time delay for snap-through, $T_\snap$, is defined as the time after $\mu=\mu_\fold$ at which the midpoint of the arch reaches its maximum (see the label in fig.~\ref{fig:RawTrajectories}c). In (a) raw results are shown, while (b) shows that the duration of snap-through is not controlled by the damping $\nu$ alone. (Note that in (a) the $x$-axis shows $|\dot{d}|$, while in (b) $|\mudot|=\dot{d}/\alpha^2$ is used.)}
\label{fig:RawSnapTimes}
\end{figure}

Repeating similar simulations with different inclination angles shows that the time\footnote{Note that we may present result in terms of either the snap-through time $T_\snap$ or the lag in the bifurcation parameter $|\dMuEff|=|\mudot| T_\snap$, as defined in fig.~\ref{fig:BifurcationsSchematic}b. We shall see that $T_\snap$ is the parameter that emerges most naturally from our analysis.} at which snap-through is observed, $T_\snap$, depends on the rate of ramping $\dot{d}$, clamp angle $\alpha$, and the viscous damping coefficient $\nu$, as shown in fig.~\ref{fig:RawSnapTimes}a. It is natural to attribute such a delay to viscous behaviour alone, so that one might expect $T_\snap$ is simply proportional to the dimensionless viscosity $\nu$. To show that this is not the case, fig.~\ref{fig:RawSnapTimes}b shows the snap-through time rescaled by $\nu$: these results do not collapse on a single curve, showing that the role of viscosity must be more subtle than expected. We therefore move on to consider an asymptotic analysis of the quasi-linear problem to illuminate this subtle behaviour.

\section{Asymptotic analysis of the quasi-linear problem}
\label{sec:Asymptotics}

Having seen in \S\ref{sec:QuasiLinStatics} that the quasi-linear approximation allows analytical predictions for the static bifurcation behaviour  --- in excellent agreement with the fully-nonlinear static problem --- we now apply it to the dynamic scenario of most interest. The essential idea of this analysis is to exploit the smallness of $|\mudot|$ (see the colourbar in fig.~\ref{fig:RawSnapTimes} for the values used in simulations) and examine the behaviour  when $|\mu-\mu_\fold| \ll 1$, during which the arch is expected to be in the neighbourhood of the fold. In this neighbourhood, the presence of the saddle-node bifurcation slows the dynamics down in a relatively long `bottleneck' phase \cite[][]{strogatz}. This analysis leads to an amplitude equation, \eqref{eigen_equation}, which can be analysed to yield approximations for the snap-through time, $T_\snap$, depending on the viscosity $\nu$ and ramping rate $|\mudot|$. We present this analysis of the amplitude equation in \S\ref{sec:AmpEqnAnalysis}, but first detail the derivation of \eqref{eigen_equation}.

\subsection{Snap-through dynamics}
To focus on the bottleneck phase of the motion (close to the snap-through transition), we rescale time as $T=|\mudot|^{-\eta}\mathcal{T}$ for some (currently unknown) exponent $\eta > 0$; the dynamic beam equation \eqref{eqn:ElasticaLinDyn}, in terms of the rescaled displacement $W(X,T) = Y(X,T)/\alpha$ and $N_X = -\tau^2$, then becomes
\begin{equation}
|\mudot|^{2\eta} \frac{\partial^2 W}{\partial \mathcal{T}^2} + \nu |\mudot|^{\eta} \frac{\partial W}{\partial \mathcal{T}} + \frac{\partial^4W}{\partial{X}^4} + {\tau}^2 \frac{\partial^2W}{\partial{X}^2}=0,\ \ \ 0< X< 1.
\label{rescaledbeamfun}
\end{equation} This is to be solved with the usual boundary conditions, which in rescaled terms read (subscripts denoting partial differentiation) 
\begin{equation}
W(0,\mathcal{T}) = 0, \quad W_X(0,\mathcal{T})=1,\quad W(1,\mathcal{T})=W_X(1,\mathcal{T})=0,
\label{rescaledboundary}
\end{equation} 
and the constraint \eqref{eqn:LinConstraint}, which becomes
\begin{equation}
\int^1_0\left(\frac{\partial W}{\partial X}\right)^2\:\upd X=2 \mu(T) =2 \left[\mu_{\fold}(\alpha)-|\mudot|^{1-\eta}\mathcal{T}\right].
\label{rescaledconstraint}
\end{equation}

We seek a solution of \eqref{rescaledbeamfun} as an asymptotic expansion  about the equilibrium solution at the fold point; we write this `fold shape' as $W_\fold(X)$, which is given by the expression in \eqref{eqn:EqmShapeLin} when $\tau=\tau_\fold$ (as given in \eqref{eqn:FoldProperties}). In fact, we seek a regular expansion in powers of $|\mudot|^\gamma$, where the exponent $\gamma > 0$ is also to be determined:
\begin{align}
W(X,\mathcal{T})\ &=\ W_{\fold}(X)+|\mudot|^{\gamma}\: W_0(X,\mathcal{T})+|\mudot|^{2\gamma}\: W_1(X,\mathcal{T})+\dots, \label{eqn:WExpansion}\\
\tau(\mathcal{T})\ &=\ \tau_{\fold}+|\mudot|^{\gamma}\: {\tau}_0(\mathcal{T})+|\mudot|^{2\gamma}\: {\tau}_1(\mathcal{T})+\dots.
\label{eqn:TauExpansion}
\end{align} 

 Inserting the expansions \eqref{eqn:WExpansion}--\eqref{eqn:TauExpansion} into \eqref{rescaledbeamfun}, we find that, since the fold shape $W_\fold(X)$ is time-independent, the time derivatives enter only at higher order. Consequently, the first non-trivial problem, which involves terms at $O(|\mudot|^{\gamma})$ in \eqref{rescaledbeamfun}, is quasi-static. We refer to this as the leading-order problem, since it involves the lowest-order perturbation $(W_0,\tau_0)$ to the fold shape in \eqref{eqn:WExpansion}--\eqref{eqn:TauExpansion}. We will show below that this problem is, in fact, an eigenvalue problem for $(W_0,\tau_0)$ for which the time dependence only enters via the solution amplitude. It is only in the first-order problem, which occurs at $O(|\mudot|^{2\gamma})$ in \eqref{rescaledbeamfun}, that time-dependent terms appear. To avoid a quasi-static problem at this order, we require a balance between the inertia term in \eqref{rescaledbeamfun} and the first-order perturbation $(W_1,\tau_1)$, which gives, upon comparing exponents of $|\mudot|$, the relation $2\eta + \gamma = 2\gamma$. Moreover, to obtain non-trivial dynamics, the ramping term must enter the end-shortening constraint at first-order, implying $2\gamma = 1-\eta$. Combining these relations gives
\begin{equation*}
\eta = \frac{1}{5}, \quad \gamma = \frac{2}{5}.
\end{equation*}
With these exponents determined, we now explicitly solve the leading and first-order problems.

\subsection{Leading-order problem $O(|\mudot|^{2/5})$}

Considering terms of $O(|\mudot|^{2/5})$ in \eqref{rescaledbeamfun}, we obtain the homogeneous linear equation:
\begin{equation}
L({W_0},{{\tau}_0})\ {\equiv}\ \frac{\partial^4{W_0}}{\partial{X}^4}+{\tau_{\fold}^2} \frac{\partial^2{W_0}}{\partial{X}^2}+2{\tau_{\fold}}{{\tau}_0}\frac{\upd^2W_{\fold}}{\upd X^2}=0.
\label{Leadingorderhomoequ}
\end{equation}\\
The boundary conditions \eqref{rescaledboundary} and end-shortening constraint \eqref{rescaledconstraint} are also homogeneous at $O(|\mudot|^{2/5})$:
$$
\int^1_0{\frac{\upd W_{\fold}}{\upd X}{\frac{\partial W_0}{\partial X}}}~\upd X=0,\quad W_0(0,\mathcal{T})=W_{0,X}(0,\mathcal{T})=W_0(1,\mathcal{T})=W_{0,X}(1,\mathcal{T})=0.
$$\\
Using linearity of the operator $L({\cdot},{\cdot})$, the parameter ${\tau}_0$ can be scaled out from \eqref{Leadingorderhomoequ}, so that
\begin{equation}
(W_0,{\tau}_0)=A(\mathcal{T})\times(W_p(X),1).
\label{Amplitudescaleout}
\end{equation}
Here $A(\mathcal{T})=\tau_0(\mathcal{T})$ is a time-dependent amplitude and $W_p(X)$ is the eigenfunction satisfying \eqref{Leadingorderhomoequ} with $\tau_0=1$, i.e.
\begin{equation}
L(W_p,1)\ {\equiv}\ \frac{\upd^4{W_p}}{\upd{X}^4}+{\tau_{\fold}^2} \frac{\upd^2W_p}{\upd X^2}+2\tau_{\fold}\frac{\upd^2W_{\fold}}{\upd X^2}=0,
\label{Leadingorderhomoequ:Resc}
\end{equation} 
subject to
\begin{equation}
\int^1_0{\frac{\upd W_{\fold}}{\upd X}{\frac{\upd W_p}{\upd X}}}~\upd X=0, \quad W_p(0)=W_{p}'(0)=W_p(1)=W_{p}'(1)=0.
\label{BoundaryWp}
\end{equation}

A unique solution\footnote{This system appears to over-determine $W_p(X)$, as there are four derivatives in \eqref{Leadingorderhomoequ:Resc} but five constraints in \eqref{BoundaryWp}. A solution is found by applying the boundary conditions in \eqref{BoundaryWp} to the solution of \eqref{Leadingorderhomoequ:Resc}, but it can then be shown that this solution also satisfies the integral constraint exactly. Specifically, this is because $W_p$ corresponds to a `neutrally-stable' or `slow' eigenfunction about the fold shape $W_\fold(X)$ (i.e.~a small-amplitude vibrational mode whose natural frequency is zero). Since the inverted shape coincides with a linearly unstable solution at the fold point, the system has a zero eigenvalue there and hence a non-trivial solution for $W_p$ exists. The choice $\tau_0 =1$ serves as a normalization condition to uniquely specify $W_p$.} of \eqref{Leadingorderhomoequ:Resc}--\eqref{BoundaryWp} can be found as
\begin{equation}
W_p(X)=\frac{1}{\tau_{\fold}}\left(X \frac{\upd W_{\fold}}{\upd X}-X\right) +a_1(\sin\tau_{\fold}X-\tau_{\fold}X)+a_2(\cos\tau_{\fold}X-1),
\label{SolutionWp}
\end{equation}
where
\begin{equation*}
a_1=-2\frac{\sin^2(\tau_{\fold}/2) [(\tau_{\fold}^2-2)\cos\tau_{\fold}-2\tau_{\fold}\sin\tau_{\fold}+2]}{\tau_{\fold}^2(2\cos\tau_{\fold}+\tau_{\fold}\sin\tau_{\fold}-2)^2},
\end{equation*}
and
\begin{equation*}
a_2=-\frac{\tau_{\fold}^3+\tau_{\fold}^2\sin\tau_{\fold}(\cos\tau_{\fold}-2)+2(\tau_{\fold}\cos\tau_{\fold}-\sin\tau_{\fold})(\cos\tau_{\fold}-1)}{\tau_{\fold}^2(2\cos\tau_{\fold}+\tau_{\fold}\sin\tau_{\fold}-2)^2}.
\end{equation*}
We shall see that what is really required from this solution is actually two integrals of $W_p(X)$, which we record here:
\begin{equation}
I_1=\int^1_0{W_p^2}~\upd X \approx 0.01630,\ \ I_2=\int^1_0{\left(\frac{\upd W_p}{\upd X}\right)^2}~\upd X \approx 0.2993.
\label{IntegralsI1I2}
\end{equation}

\subsection{First-order problem $O(|\mudot|^{4/5})$}

Continuing the expansion of \eqref{rescaledbeamfun} to $O(|\mudot|^{4/5})$, and using \eqref{Amplitudescaleout} to eliminate $W_0$ for the amplitude $A$, we find that
\begin{equation}
L(W_1,\tau_1)=- \left(\frac{\upd^2A}{\upd \mathcal{T}^2} + \nu |\mudot|^{-1/5} \frac{\upd A}{\upd \mathcal{T}} \right)W_p -A^2 \left( 2\tau_{\fold}\frac{\upd^2W_p}{\upd X^2} + \frac{\upd^2W_{\fold}}{\upd X^2} \right),
\label{EigenfunW1tau1}
\end{equation}
where the operator $L(\cdot,\cdot)$ is as defined in \eqref{Leadingorderhomoequ}. The end-shortening constraint \eqref{rescaledconstraint} at $O(|\mudot|^{4/5})$ becomes
\begin{equation}
\int^1_0{\frac{\upd W_{\fold}}{\upd X}{\frac{\partial W_1}{\partial X}}}~\upd X=-\frac{I_2}{2} A^2-\mathcal{T},
\label{Int_Wfx&W1x}
\end{equation}
while the boundary conditions  \eqref{rescaledboundary} are
$$
W_1(0,\mathcal{T})=W_{1,X}(0,\mathcal{T})=W_1(1,\mathcal{T})=W_{1,X}(1,\mathcal{T})=0.
$$

The inhomogeneous equation \eqref{EigenfunW1tau1} features the same linear operator $L({\cdot},{\cdot})$ that was defined in \eqref{Leadingorderhomoequ}, having arisen in the leading-order problem. According to the Fredholm Alternative Theorem, a solution of \eqref{EigenfunW1tau1} can exist only if the right-hand side satisfies a solvability condition \citep{Keener}. We now show how this solvability condition leads to an equation for the evolution of $A(\mathcal{T})=\tau_0(\mathcal{T})$ --- the amplitude equation.

\subsection{Amplitude equation for $A(\mathcal{T})$}

To proceed, we note that \eqref{EigenfunW1tau1} may be written $L(W_1,\tau_1)=f(X;A)$ --- the right-hand side is a known function of $X$ with time-dependence entering only via the function $A(\mathcal{T})$. Multiplying by $W_p(X)$ and integrating over the domain $[0,1]$ gives 
\begin{equation}
\int^1_0L(W_1,\tau_1)\: W_p~\upd X=\int^1_0f \cdot \: W_p~\upd X.
\label{IntMultply_LW1=fA}
\end{equation}
We proceed to evaluate the left-hand side of \eqref{IntMultply_LW1=fA} by repeated integration by parts to give
\beq
\int_0^1L(W_1,\tau_1) \: W_p~\upd X=\int_0^1\left(\frac{\upd^4W_p}{\upd X^4}+\tau^2_\fold\frac{\upd^2W_p}{\upd X^2}\right) \: W_1~\upd X.
\eeq 
After using \eqref{Leadingorderhomoequ:Resc}, this gives
\beq
\int_0^1L(W_1,\tau_1) \: W_p~\upd X=-2\tau_\fold\int_0^1\frac{\upd^2W_\fold}{\upd X^2}\: W_1~\upd X=-2\tau_\fold\left(\frac{I_2}{2}A^2+\mathcal{T}\right),
\label{eqn:LfirstInt}
\eeq where in the last equality we have used \eqref{Int_Wfx&W1x} to eliminate $\int_0^1W'_\fold\partial W_1/\partial X~\upd X$. To progress further, we return to the right-hand side of  \eqref{IntMultply_LW1=fA}, which can be expressed as
\begin{equation}
\int^1_0f \cdot \: W_p~\upd X= \int^1_0\left[-W_p \left(\frac{\upd^2{A}}{\upd \mathcal{T}^2} + \nu |\mudot|^{-1/5} \frac{\upd{A}}{\upd \mathcal{T}} \right) - A^2 \left( 2{\tau_{\fold}}\frac{\upd^2{W_p}}{\upd X^2} + \frac{\upd^2{W_{\fold}}}{\upd X^2} \right)\right]W_p~\upd X,
\label{Int_fA&Wp}
\end{equation} leading to (using \eqref{IntegralsI1I2}):
\begin{equation}
\int^1_0 f \cdot\: W_p~\upd X= -\left(\frac{\upd^2{A}}{\upd \mathcal{T}^2} + \nu |\mudot|^{-1/5} \frac{\upd A}{\upd \mathcal{T}} \right) I_1 + 2{\tau_{\fold}} I_2 A^2. 
\label{Inted_fA&Wp}
\end{equation}
Substituting equations \eqref{eqn:LfirstInt} and \eqref{Inted_fA&Wp} into \eqref{IntMultply_LW1=fA}, we immediately obtain
\begin{equation}
\frac{\upd^2{A}}{\upd \mathcal{T}^2} + \nu |\mudot|^{-1/5} \frac{\upd A}{\upd \mathcal{T}} = C_1 \mathcal{T} + C_2 A^2,
\label{eigen_equation}
\end{equation}
where 
\begin{equation}
C_1 = \frac{2\tau_\fold}{I_1} \approx 930.6, \quad C_2 = \frac{3\tau_{\fold} I_2}{I_1} \approx 417.8.
\label{value_c1_c2}
\end{equation}
Finally, to remove the numerical pre-factors in \eqref{eigen_equation}, we make one further substitution, letting $\tbar=C_1^{1/5}C_2^{1/5}\mathcal{T}$ and $\A=C_1^{-2/5}C_2^{3/5}A$, to obtain
\beq
\frac{\upd^2\A}{\upd \tbar^2} + \Lambda \frac{\upd\A}{\upd \tbar} = \tbar + \A^2
\label{EigenEqn:Alpha},
\eeq where
\beq
\Lambda=(C_1C_2)^{-1/5} \: \nu|\mudot|^{-1/5}.
\label{eqn:LambdaDefn}
\eeq

The amplitude equation \eqref{EigenEqn:Alpha} represents a great simplification of the original problem we began with, which comprised the partial differential equation \eqref{rescaledbeamfun} subject to the nonlinear, non-local constraint \eqref{rescaledconstraint} for the applied end-shortening. The amplitude equation \eqref{EigenEqn:Alpha} has the canonical form of dynamics close to a saddle-node bifurcation \cite[][]{strogatz} in  which the bifurcation parameter is ramped linearly in time. We see that both first and second-order time derivatives are present: both viscous damping and beam inertia are present in the problem. Interestingly, the precise form of the boundary conditions applied to the arch enters only through the dimensionless constants $C_1$ and $C_2$ defined in \eqref{value_c1_c2} (the boundary conditions determine the functions $W_p(X)$ and $W_{\fold}(X)$, and hence the values of the integrals $I_1$ and $I_2$ in \eqref{IntegralsI1I2}). Even then, these constants were scaled out in \eqref{EigenEqn:Alpha} and so we expect \eqref{EigenEqn:Alpha} to be generic for snap-through instabilities in which the relevant control parameter is ramped through a saddle-node bifurcation. 

A key feature of \eqref{EigenEqn:Alpha} is that the damping coefficient $\nu$ only enters via the combination $\nu|\mudot|^{-1/5}$ that appears in the dimensionless parameter $\Lambda$. We therefore anticipate two limiting regimes: in the first, observed for sufficiently large $\nu$ (or small $|\mudot|$), we expect viscous damping to dominate the inertial term (and the dynamics in the bottleneck region to be overdamped to leading order). However, a second regime exists: for sufficiently small $\nu$ compared to $|\mudot|$, inertial forces will instead dominate --- the bottleneck dynamics are then limited by how quickly the beam can be accelerated. We will make these considerations more precise, and quantify what happens in each case, in \S\ref{sec:AmpEqnAnalysis}.

\section{Analysis of the amplitude equation \label{sec:AmpEqnAnalysis}}

Another important feature of the amplitude equation \eqref{EigenEqn:Alpha} is that solutions exhibit finite-time blow-up, i.e.~$\A\to\infty$ at some finite time $\tbar=\tbar_\infty$. This is caused by the quadratic forcing term in \eqref{EigenEqn:Alpha}. In fact, as $\A\to\infty$ and the motions become increasingly rapid, the equation must reduce to a balance between inertia and the quadratic forcing term, which admits a power-law solution of the form
\beq
\A \sim 6 \left(\tbar_\infty - \tbar \right)^{-2}, \label{eqn:PowerLawBlowUp}
\eeq
where $\tbar_\infty < \infty$  is the blow-up time. Since the asymptotic analysis of \S\ref{sec:Asymptotics} applies only in the neighbourhood of the fold, this analysis must break down as this finite-time blow-up is approached. (Formally, from the definition of the  amplitude $A$ in \eqref{Amplitudescaleout}, we see that asymptotic validity of the expansions in \eqref{eqn:WExpansion}--\eqref{eqn:TauExpansion} is lost once $A$, and hence the rescaled amplitude $\A$, reaches $O(|\mudot|^{-2/5}) \gg 1$.) Nevertheless, we use the blow-up time as an approximation of the snap-through time: the power law solution for $\A$ in \eqref{eqn:PowerLawBlowUp} implies that the break down of the asymptotic analysis occurs when
$\tbar_\infty - \tbar = O(|\mudot|^{1/5}) \ll 1$, i.e.~when the rescaled time is $\tbar_\infty$ to leading order in $|\mudot|^{1/5}$. As the duration of snap-through is dominated by the time during which the system remains in the bottleneck, we expect the total snapping time to equal the blow-up time to leading order, i.e.
\beq
\Tsnap =\tbar_\infty + O(|\dot{\mu}|^{1/5}).
\eeq

We now turn to the calculation of $\tbar_\infty$, which is, in general, a function of the rescaled damping coefficient $\Lambda$. To determine $\tbar_\infty(\Lambda)$ we seek to solve the amplitude equation \eqref{EigenEqn:Alpha}, but to do so requires first a careful discussion of the appropriate matching conditions for the solution.

\subsection{Matching conditions for $\A$}

To solve the amplitude equation \eqref{EigenEqn:Alpha} requires appropriate initial conditions for $\A$ and $\dot{\A}$. It is tempting to assume that these are homogeneous, i.e.~$\A(0)=\dot{\A}(0)=0$ --- similar initial conditions were used in other snap-through problems, in which the system starts at rest in the vicinity of the saddle-node bifurcation with the control parameter fixed \citep{gomez2017critical,gomezthesis}. In the problem considered here, however, the bifurcation parameter is ramped and the system starts on the stable branch of equilbrium solutions well before the saddle-node bifurcation is approached. We therefore need to ensure the solution matches back onto this branch at large, negative times. This is equivalent to requiring
\beq
\A \sim -(-\tbar)^{1/2} \quad \mathrm{as} \quad \tbar \to -\infty.
\label{eqn:MatchingDefn}
\eeq

For general $\Lambda$ we are unable to solve \eqref{EigenEqn:Alpha} analytically, and so will make extensive use of numerical solutions. To enforce the matching condition \eqref{eqn:MatchingDefn} numerically, we first determine the series expansion of the solution about $\tbar=-\infty$; we expand in powers of $(-\tbar)^{-1/2}$ with the leading order term as above (or alternatively using the change of variables $u=1/(-\tbar)$ and expanding in powers of $u^{1/2}$). Substituting this expansion into \eqref{EigenEqn:Alpha} and solving at successive orders, we obtain
\begin{eqnarray}
\A & = & -(-\tbar)^{1/2} - \frac{\Lambda}{4}(-\tbar)^{-1} - \frac{1}{8}(-\tbar)^{-2} + \frac{5\Lambda^2}{32}(-\tbar)^{-5/2} + \frac{13\Lambda}{32}(-\tbar)^{-7/2} \nonumber \\
&& \: - \frac{15\Lambda^3}{64}(-\tbar)^{-4} + \frac{49}{128}(-\tbar)^{-9/2} +O(-\tbar)^{-5}.
\label{eqn:alphaAsy}
\end{eqnarray}
If $\Lambda \lesssim O(1)$, we see that the perturbation to $-(-\tbar)^{1/2}$ is indeed small provided $(-\tbar) \gg 1$. If $\Lambda \gg 1$, however, the first-order term becomes comparable to the leading-order term if $(-\tbar)$ is not sufficiently large. Later, by analysing in detail the early-time behaviour when $\Lambda \gg 1$, we will show that the expansion is still valid provided the leading-order term dominates the first-order term, i.e.,~
\beq
(-\tbar) \gg \Lambda^{2/3}.
\eeq
The expansion can then be used to obtain initial conditions for numerical solutions of \eqref{EigenEqn:Alpha} that start at some large, negative time $\tbar_{\mathrm{start}}$ with $(-\tbar_{\mathrm{start}}) \gg 1$ and $(-\tbar_{\mathrm{start}}) \gg \Lambda^{2/3}$. In the numerical solutions reported below, we take
\beq
\tbar_{\mathrm{start}}=-100\times\max\{1,\Lambda^{2/3}\},
\label{eqn:tstart}
\eeq
and integrate in \textsc{matlab} using the routine \texttt{ode15s} (relative and absolute error tolerances of size $10^{-10}$).

Before discussing the snap-through time for general $\Lambda$, however, we analyse the solution in the limiting cases $\Lambda \ll 1$ and $\Lambda \gg 1$.

\subsection{Underdamped snap-through: $\Lambda \ll 1$}

We first consider the underdamped limit $\Lambda \ll 1$, for which \eqref{EigenEqn:Alpha} reduces to
\beq
\frac{\upd^2\A}{\upd \tbar^2} \sim \tbar + \A^2.
\label{EigenEqn:AlphaSmallLam}
\eeq 
The matching condition \eqref{eqn:alphaAsy} at leading order becomes
\beq
\A \sim -(-\tbar)^{1/2} - \frac{1}{8}(-\tbar)^{-2} + \frac{49}{128}(-\tbar)^{-9/2} \quad \mathrm{as} \quad \tbar \to -\infty.
\label{eqn:alphaAsySmallLam}
\eeq
In contrast to the case when the bifurcation parameter is fixed during snap-through \citep{gomez2017critical}, the non-autonomous ramping term prevents us from integrating \eqref{EigenEqn:AlphaSmallLam} analytically. Numerically integrating \eqref{EigenEqn:AlphaSmallLam}, we find that blow-up occurs at
\beq
\tbar_{\infty} \approx 3.4039.
\eeq
Recalling our approximation that $\Tsnap\approx\tbar_\infty$, and undoing the various rescalings of the time variable performed so far (defined just before equations \eqref{rescaledbeamfun} and \eqref{EigenEqn:Alpha}), we obtain the following approximation for the snap-through time expressed in the original dimensionless time $T$ used in the elastica simulations:
\beq
T_{\snap} \approx |\mudot|^{-1/5}(C_1 C_2)^{-1/5}\: \Tsnap \approx 0.2594\:|\mudot|^{-1/5}.
\label{eqn:SmallLam}
\eeq Alternatively, the delay in the parameter value at which snap-through is observed, $|\dMuEff|=|\mudot|T_\snap$, is given by
\begin{equation}
|\dMuEff|=|\mudot|\: T_\snap\approx0.2594\:|\mudot|^{4/5}.
\end{equation}

\begin{figure}
\centering
\includegraphics[width=0.9\textwidth]{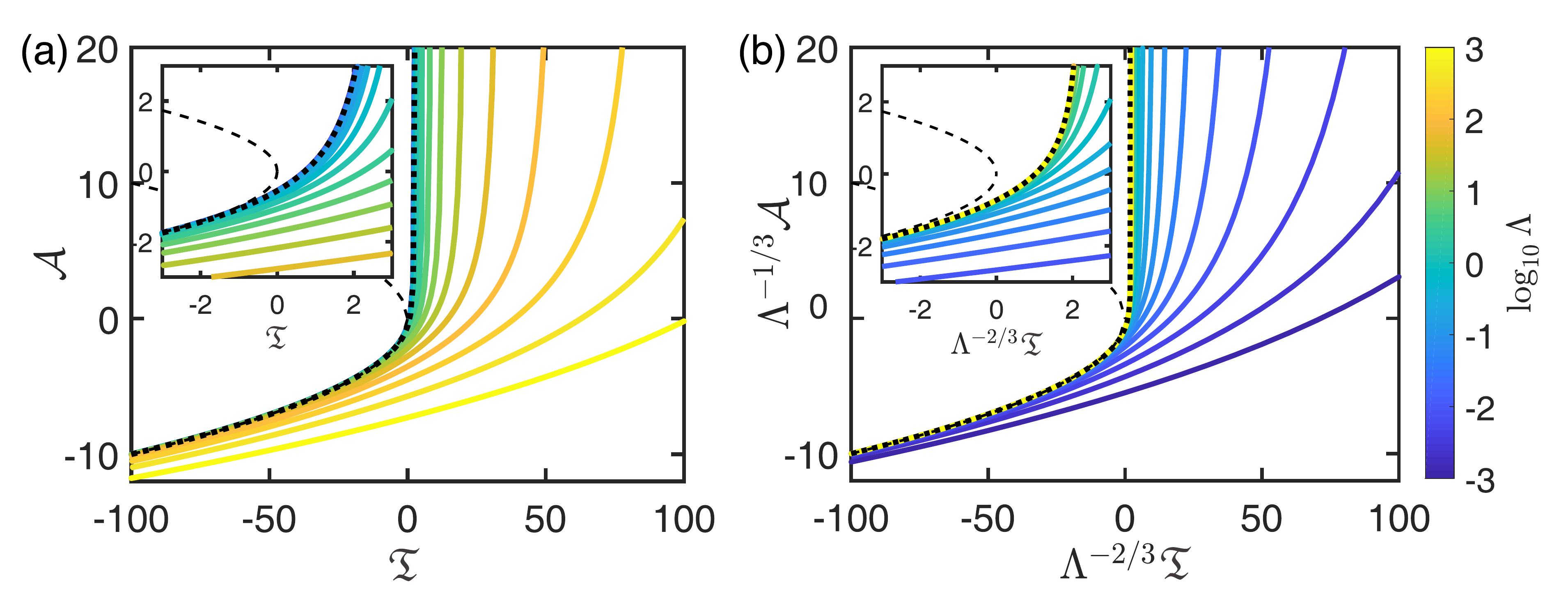}
\caption{Dynamics in the bottleneck phase. (a) Trajectories $\A(\tbar)$ obtained by numerically integrating the amplitude equation \eqref{EigenEqn:Alpha} with matching condition \eqref{eqn:alphaAsy} for fixed $\Lambda$ (coloured curves; see colourbar, right). (The matching condition ensures that for large, negative times the solution approaches the stable branch of the bifurcation diagram, which is shown by the black long-dashed curve.) Also shown is the limiting solution with $\Lambda = 0$ (black dotted curve). The inset displays a close-up of the behaviour near the fold point, $(0,0)$. (b) Rescaling dimensionless time and the amplitude variable to balance all terms in the overdamped limit \eqref{eqn:EigenOD}, the trajectories for large $\Lambda$ collapse onto the analytical solution \eqref{eqn:AirySoln} (black dotted curve). The inset again displays a close-up of the behaviour near the fold point.}
\label{fig:BottleneckTrajs}
\end{figure}

The trajectory obtained by integrating \eqref{EigenEqn:AlphaSmallLam} numerically with matching condition \eqref{eqn:alphaAsySmallLam} is plotted as a black dotted curve on fig.~\ref{fig:BottleneckTrajs}a. (For later reference, trajectories obtained by integrating the full amplitude equation \eqref{EigenEqn:Alpha} for different $\Lambda$ are also shown.) From the inset in fig.~\ref{fig:BottleneckTrajs}a, the `lag' of the system behind the quasi-static solution branch (black long-dashed curve) near the fold point is clearly visible. In addition to the snap-through time, we can quantify this lag in two ways: firstly, through the value of the amplitude at $\tbar=0$, i.e.~$\A(0)$. Recalling the definition of the amplitude variable $A$ in \eqref{Amplitudescaleout}, $\A(0)$ measures the size of the leading-order perturbation from the fold shape $(W_{\fold},\tau_{\fold})$ when $\tbar = 0$ or equivalently $\mu = \mu_{\fold}$. Secondly, we may measure the time at which this amplitude $\A$ crosses zero, which we label $\tbar\lvert_{\A = 0} > 0$. This  corresponds to the time at which the leading-order shape perturbation vanishes and so the shape is closest to the  shape at the fold, given by \eqref{eqn:EqmShapeLin} with $\tau=\tau_\fold$. Using the numerical solution for $\Lambda = 0$, we evaluate these quantities as
\beq
\A(0) \approx -0.5496, \quad \tbar\lvert_{\A = 0} \approx 0.7154.
\eeq
Using \eqref{eqn:TauExpansion} and \eqref{Amplitudescaleout} to write $\A$ back in terms of the compressive force $\tau$, this gives
\begin{equation*}
\tau\lvert_{\mu = \mu_{\fold}} \sim \tau_{\fold} + |\mudot|^{2/5}C_1^{2/5}C_2^{-3/5}\A(0) \approx \tau_{\fold} - 0.2264\:|\mudot|^{2/5},
\end{equation*}and
\begin{equation*}
T\lvert_{\tau = \tau_{\fold}} \sim  |\mudot|^{-1/5}(C_1 C_2)^{-1/5} \tbar\lvert_{\A = 0} \approx 0.0545\:|\mudot|^{-1/5}.
\end{equation*} Note that the latter result shows that the time-lag for the arch to approach its fold shape \emph{diverges} as the ramping rate $|\mudot|\to0$. 

\subsection{Overdamped snap-through: $\Lambda\gg1$}

In the opposite limit of very large damping, $\Lambda\gg1$, we expect the inertial term in \eqref{EigenEqn:Alpha}, $\upd^2\A/\upd\tbar^2$, to be negligible and hence that
\beq
\Lambda \frac{\upd \A}{\upd \tbar} \sim \tbar + \A^2.
\label{eqn:EigenOD}
\eeq 
(This is mathematically equivalent to introducing a further rescaling such that $\tbar = O(\Lambda^{2/3})$ and $\A = O(\Lambda^{1/3})$, which balances the first-order derivative in \eqref{EigenEqn:Alpha} with both terms on the right-hand side; we then neglect the inertial term, which would become a factor $\Lambda^{-5/3} \ll 1$ smaller in these new variables.) Upon making the change of variables \cite[see, for example, Appendix B of][]{erneux1989}
\begin{equation*}
\tbar = -\Lambda^{2/3}\mathscr{T}, \quad \A = \Lambda^{1/3} \frac{1}{\phi}\frac{\upd \phi}{\upd \mathscr{T}},
\end{equation*}
equation \eqref{eqn:EigenOD} transforms to the Airy equation for $\phi(\mathscr{T})$:
\begin{equation*}
\frac{\upd^2 \phi}{\upd \mathscr{T}^2} \sim \mathscr{T}\phi. 
\end{equation*}
 The matching condition $\A\sim-(-\tbar)^{1/2}$ implies that $\phi$ decays exponentially for large arguments, so the solution is proportional to the Airy function of the first kind, i.e.~$\phi \propto \Ai(\mathscr{T})$; determining the coefficient and transforming back gives
\beq
\A \sim \Lambda^{1/3}\frac{\Ai'(-\Lambda^{-2/3}\:\tbar)}{\Ai(-\Lambda^{-2/3}\:\tbar)},
\label{eqn:AirySoln}
\eeq which may readily be seen to be a solution of \eqref{eqn:EigenOD}\footnote{Expanding this solution for $(-\Lambda^{-2/3}\:\tbar) \gg 1$, we recover the series \eqref{eqn:alphaAsy} found earlier for general $\Lambda$ after rescaling $\tbar = O(\Lambda^{2/3})$, $\A = O(\Lambda^{1/3})$ and neglecting terms of $O(\Lambda^{-5/3}) \ll 1$; this justifies our earlier assumption that when $\Lambda \gg 1$, the expansion \eqref{eqn:alphaAsy} is still valid provided we are at sufficiently early times $(-\tbar) \gg \Lambda^{2/3}$.}.  We note that similar Airy function solutions have been obtained in several other problems involving overdamped dynamics near a saddle-node bifurcation, in which the control parameter is ramped linearly in time \citep{erneux1989,laplante1991,zhu2015,majumdar2013,li2019}.

In this overdamped limit, the finite-time singularity occurs at $-\Lambda^{-2/3}\:\tbar_\infty \sim \mathscr{T}_\ast$ where $\mathscr{T}_\ast\approx-2.3381$ is the root of $\Ai(\mathscr{T})=0$ that is smallest in magnitude. However, the solution \eqref{eqn:AirySoln} is not uniformly valid. In particular, expanding  \eqref{eqn:AirySoln} for $A \gg 1$, we obtain the power-law behaviour:
\beq
\A \sim \Lambda \left(-\Lambda^{2/3} \mathscr{T}_\ast -\tbar\right)^{-1}. 
\eeq
This implies that the neglected inertia term first becomes comparable to the damping term when
\beq
-\Lambda^{2/3} \mathscr{T}_\ast - \tbar = O(\Lambda^{-1}), \quad \A = O(\Lambda^2),
\eeq
after which \eqref{eqn:AirySoln} is no longer valid. Further analysis shows that the solution of the full amplitude equation \eqref{EigenEqn:Alpha} always blows up inside the interval $-\Lambda^{2/3} \mathscr{T}_\ast - \tbar = O(\Lambda^{-1})$, ultimately reducing to purely inertial dynamics with the power law behaviour \eqref{eqn:PowerLawBlowUp} that was found earlier. Recalling that the solution leaves the bottleneck phase when $\A = O(|\mudot|^{-2/5}) \gg 1$, the dynamics can therefore be underdamped or overdamped by the end of the bottleneck depending on whether $|\mudot|^{-2/5} \gg \Lambda^2$ or $|\mudot|^{-2/5} \ll \Lambda^2$, which is equivalent to $\nu \ll 1$ or $\nu \gg 1$, respectively. In both cases, we have that $\A = O(|\mudot|^{-2/5})$ when $\tbar \sim -\Lambda^{2/3} \mathscr{T}_\ast$ with error $O(\Lambda^{-1}) \ll 1$. We deduce that the bottleneck time (and hence the snap-through time) to leading order is simply the blow-up time of the Airy function solution, i.e.~
 \beq
\tbar_{\mathrm{snap}} \approx \tbar_{\infty} \sim -\Lambda^{2/3} \mathscr{T}_\ast.
\eeq
In terms of original dimensionless time $T$, this implies the snap-through time
\beq
T_\snap \approx -\mathscr{T}_\ast(C_1C_2)^{-1/5}|\mudot|^{-1/5}\Lambda^{2/3}\approx 0.1782\:|\mudot|^{-1/5}\Lambda^{2/3}.
\label{eqn:LargeLam}
\eeq 
(Note that with a fixed dimensionless viscosity $\nu$, $T_\snap\propto|\mudot|^{-1/3}$ because  $\Lambda = (C_1C_2)^{-1/5} \nu|\mudot|^{-1/5}$. For comparison with numerical results, however, it is convenient to retain the dependence on the rescaled damping parameter $\Lambda$ here.) Alternatively, the delay in the parameter value at which snap-through is observed, $|\dMuEff|=|\mudot|T_\snap$, is then
\begin{equation}
|\dMuEff|\approx 0.1782 |\mudot|^{4/5}\Lambda^{2/3}\propto |\mudot|^{2/3}\nu^{1/3}.
\end{equation} 

The Airy function solution \eqref{eqn:AirySoln} is plotted as a black dotted curve in fig.~\ref{fig:BottleneckTrajs}b, where we have rescaled variables according to \eqref{eqn:AirySoln}. (For later reference, the trajectories obtained by solving the full equation \eqref{EigenEqn:Alpha} are also shown.) As with the underdamped limit, the lag of the trajectories from the quasi-static branch near the fold point can be clearly observed (inset). From \eqref{eqn:AirySoln}, we calculate the corresponding quantities in the overdamped limit:
\beq
\A(0) \sim \frac{\Ai'(0)}{\Ai(0)}\Lambda^{1/3} \approx -0.7290\:\Lambda^{1/3}, \quad \tbar\lvert_{\A = 0} \approx 1.0188\:\Lambda^{2/3},
\eeq 
or, alternatively
\begin{eqnarray*}
\tau\lvert_{\mu = \mu_{\fold}} & \approx & \tau_{\fold} -0.3004\:|\mudot|^{2/5}\Lambda^{1/3}, \\
T\lvert_{\tau = \tau_{\fold}} & \approx & 0.0776\:|\mudot|^{-1/5}\Lambda^{2/3}.
\end{eqnarray*}

\subsection{General $\Lambda$}

For general $\Lambda$ it is not possible to make analytical progress with the full amplitude equation \eqref{EigenEqn:Alpha}, so we instead solve \eqref{EigenEqn:Alpha} numerically. We integrated  \eqref{EigenEqn:Alpha} with matching condition \eqref{eqn:alphaAsy} and twenty different values of $\Lambda\in\left[10^{-3},10^3\right]$ (chosen to be equally spaced on a logarithmic scale). The resulting trajectories $\A(\tbar)$ are plotted as solid curves in the main panel of fig.~\ref{fig:BottleneckTrajs}a; trajectories with moderately small values of $\Lambda \lesssim 10^{-1}$ collapse well onto the limiting solution with $\Lambda = 0$ (black-dotted curve), being indistinguishable on the figure. To illuminate the dynamics in the overdamped limit, the numerically-determined trajectories from fig.~\ref{fig:BottleneckTrajs}a can be rescaled in the manner appropriate for the overdamped limit \eqref{eqn:EigenOD}; see fig.~\ref{fig:BottleneckTrajs}b. In this case, we observe an excellent collapse onto the Airy function solution \eqref{eqn:AirySoln} for moderately large values $\Lambda \gtrsim 10$. Regardless of the value of $\Lambda$, the trajectories shown in fig.~\ref{fig:BottleneckTrajs}  exhibit finite-time blow-up of $\A(\tbar)$ --- this blow-up time, $\tbar_\infty$, and its dependence on $\Lambda$ are key quantities of interest.

To determine the blow-up time in the numerical solution of \eqref{EigenEqn:Alpha}, we use event detection to stop integration once $\A$ reaches some large positive value $\A_{\mathrm{stop}} \gg 1$; using the power law solution \eqref{eqn:PowerLawBlowUp}, the corresponding time, labelled $\tbar_{\mathrm{stop}}$, satisfies
\beq
\tbar_\infty - \tbar_{\mathrm{stop}} \sim \left(\frac{\A_{\mathrm{stop}}}{6}\right)^{-1/2}.
\eeq In what follows we take $\A_{\mathrm{stop}} = 10^5$, which gives  $(\A_{\mathrm{stop}}/6)^{-1/2} \approx 7.7 \times 10^{-3}$, and hence the blow-up time is determined with an error of less than $1\%$.

\begin{figure}
\centering
\includegraphics[width=0.7\textwidth]{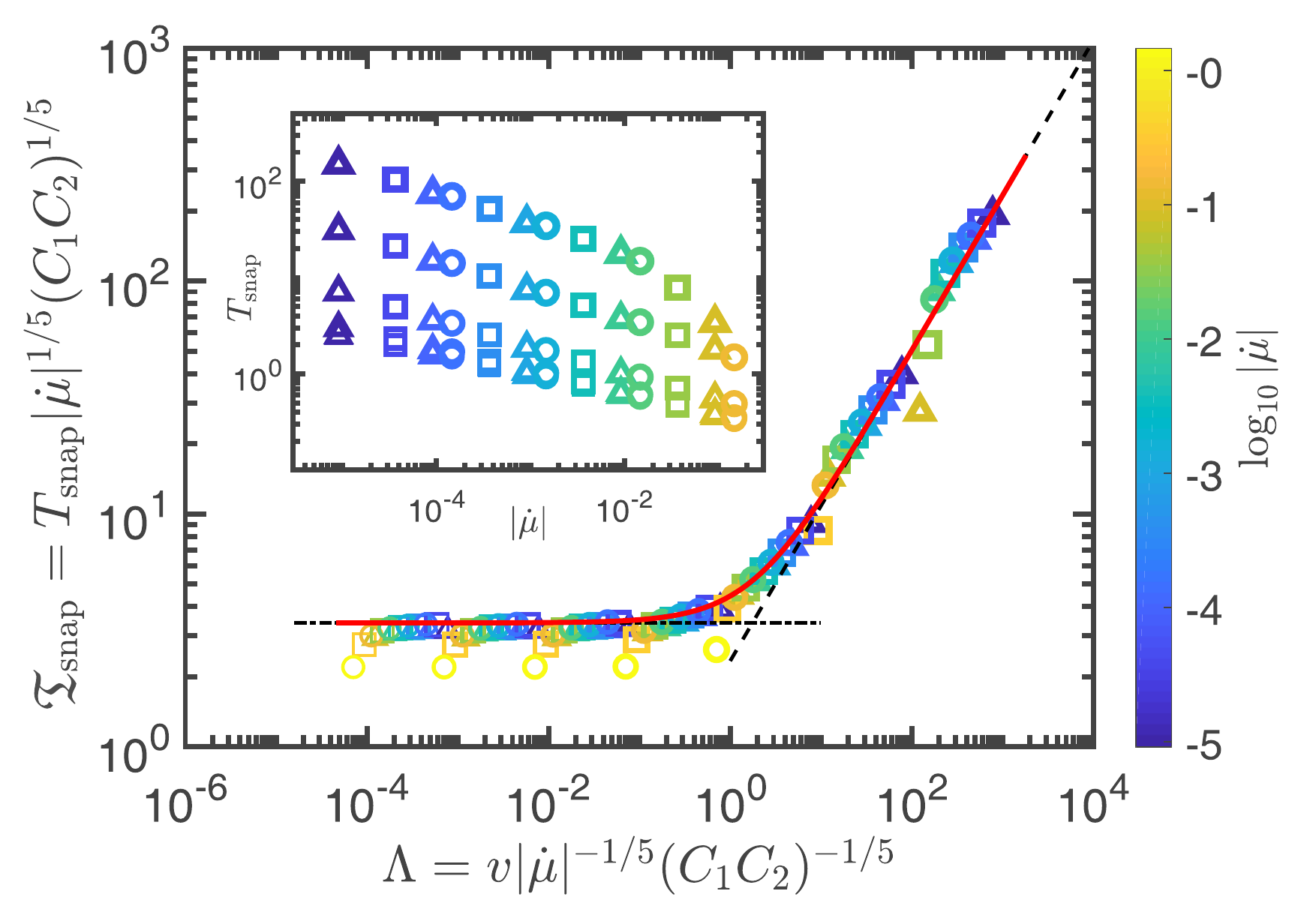}
\caption{Duration of snap-through as a function of the damping ratio $\Lambda$, as determined from the numerical simulations of the fully-nonlinear dynamic  problem (points) and the prediction of the quasi-linear model  (solid curve). Data from the numerical simulations are shown with inclination angle $\alpha$ indicated by shape (circles: $\alpha=\pi/12$, squares: $\alpha=\pi/6$, and triangles: $\alpha=\pi/3$); dimensionless dissipation $\nu$ represented by point thickness; and the normalized ramping rate, $|\mudot|$, indicated by colour (see colourbar, right). The solid curve shows the  results of integrating the amplitude equation \eqref{EigenEqn:Alpha} subject to the matching conditions \eqref{eqn:alphaAsy}, while the dashed lines show the asymptotic results for $\Lambda\ll1$ and $\Lambda\gg1$, i.e.~\eqref{eqn:SmallLam} and \eqref{eqn:LargeLam}, respectively. (Note that these data are those presented in fig.~\ref{fig:RawSnapTimes}a, and in the inset for convenience, but are rescaled according to the predictions from the quasi-linear theory.) }
\label{fig:snaptimesvsLambda}
\end{figure}

We have already seen that using the approximation $\Tsnap\approx\tbar_\infty$ introduces an error $O(|\dot{\mu}|^{1/5})$, and so the total snapping time can be approximated by
\beq
\Tsnap =\tbar_{\mathrm{stop}} + O(|\dot{\mu}|^{1/5},\A_{\mathrm{stop}}^{-1/2}).
\eeq
The dependence of $\Tsnap$ on $\Lambda$ is plotted on logarithmic axes in fig.~\ref{fig:snaptimesvsLambda} (red solid curve), and for comparison the leading-order predictions \eqref{eqn:SmallLam} and \eqref{eqn:LargeLam} in the underdamped/overdamped limits are shown (black-dotted lines). (For later comparison in \S\ref{sec:compareelastica}, numerical results from the elastica simulations are also included on the figure.) We conclude that the asymptotic analysis captures the behaviour of the amplitude equation \eqref{EigenEqn:Alpha}, both in terms of the blow-up time and the trajectory, extremely well for $\Lambda \lesssim 10^{-1}$ and $\Lambda \gtrsim 10$.

\section{Comparison with elastica simulations}
\label{sec:compareelastica}

Having understood the predictions of the amplitude equation \eqref{EigenEqn:Alpha} as a reduced model of the dynamics of snap-through, we now turn to compare these predictions with the  numerical simulations of the fully-nonlinear (elastica) dynamics of an elastic arch subject to a time-varying end-end displacement.

A first comparison is to consider the dependence of the total snap-through time on the rescaled damping coefficient $\Lambda$, defined in \eqref{eqn:LambdaDefn}. The dimensionless snap-through times from the full numerics were presented in fig.~\ref{fig:RawSnapTimes} for different values of the damping, $\nu$, and ramping rate, $\mudot$. The asymptotic analysis of \S\ref{sec:Asymptotics} and \S\ref{sec:AmpEqnAnalysis} suggests that these raw numerical data should collapse (for $|\mudot|\ll1$) when the rescaled snap-through time $\Tsnap=T_\snap|\mudot|^{1/5}(C_1C_2)^{1/5}$ is plotted as a function of $\Lambda$. Figure \ref{fig:snaptimesvsLambda} shows this rescaled data in comparison with the predictions of the asymptotic model; on the whole, we observe good collapse, and good agreement with the predictions of the asymptotic model. As should be expected, the agreement breaks down as $|\mudot|$ increases (moving from bluer to yellower points in fig.~\ref{fig:snaptimesvsLambda}) since the asymptotic analysis relies on the assumption that $|\mudot|\ll1$. Similarly, numerical results with larger values of $\alpha$ tend to deviate more from the theoretical prediction (for a given value of $|\mudot|$); this is to be expected since our derivation of the quasi-linear model relies on the assumption that $\alpha\ll1$.

\begin{figure}
\centering
\includegraphics[width=0.8\textwidth]{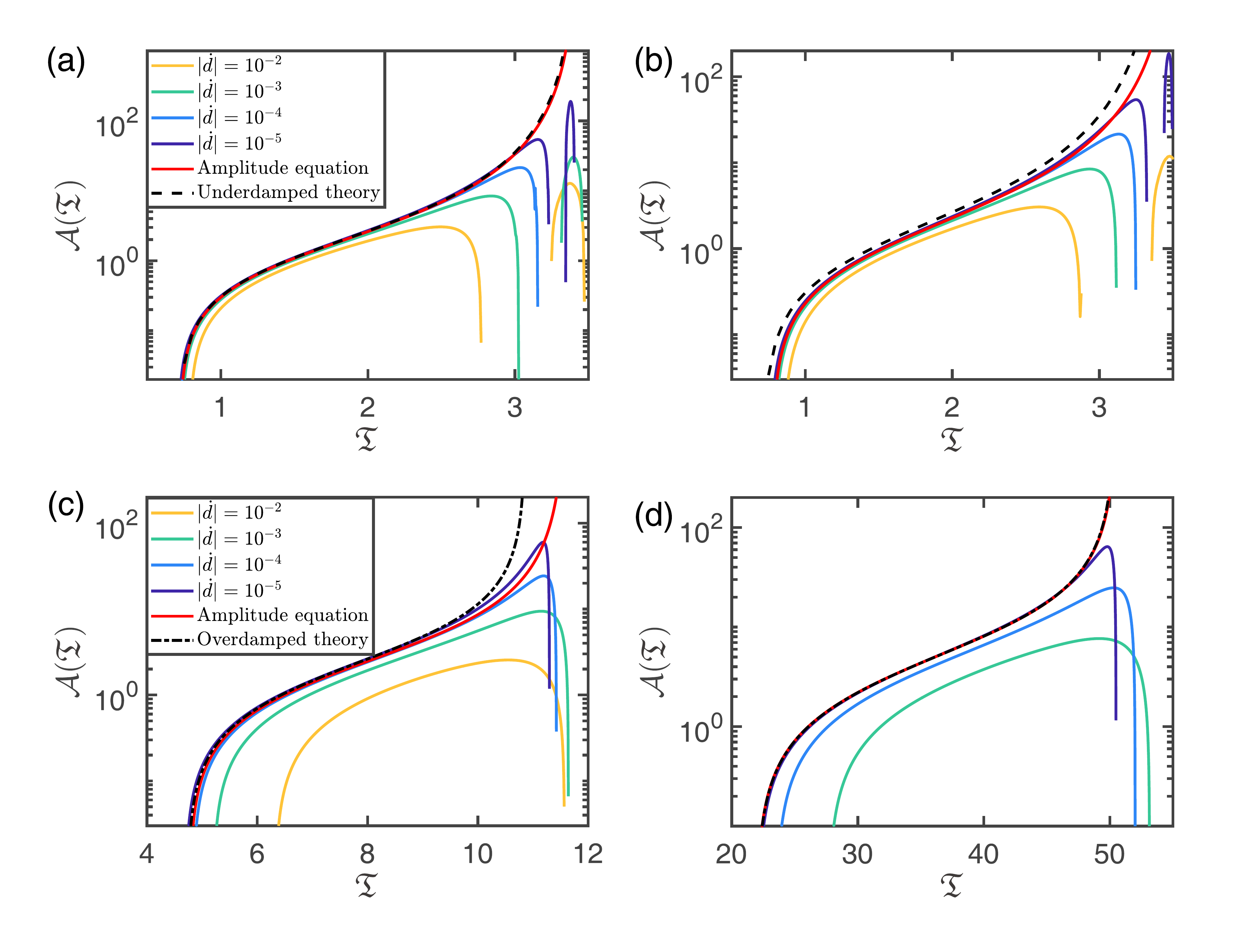}
\caption{Trajectories of the amplitude $\A(\tbar)$ determined from numerical simulations of the fully-nonlinear problem (coloured curves)  recover the trajectories predicted by the quasi-linear model (dashed black and solid red curves) as $|\dot{d}|\to0$. Results are shown for different values of the rescaled damping coefficient $\Lambda$: (a) $\Lambda = 10^{-2}$, (b) $\Lambda = 10^{-1}$, (c) $\Lambda = 10^{1}$ and (d) $\Lambda = 10^{2}$, all with $\alpha = \pi/6$. Values of the loading rate $\dot{d}$ are indicated in the legend of (a) and (c). The numerical solution of the underdamped problem, \eqref{EigenEqn:AlphaSmallLam} is shown by the dashed curve in (a) and (b) while the analytical solution for the overdamped problem, \eqref{eqn:AirySoln}, is shown by the dash-dotted curve in (c) and (d).}
\label{fig:TrajectoriesFullNums}
\end{figure}

A more stringent comparison between the asymptotic and numerical results is to compare the trajectory of the amplitude, $\A(\tbar)$, as determined from the numerical results and the asymptotic analysis. To perform such a comparison requires determination of $\A(\tbar)$ from the full numerical simulations; we invert the leading-order asymptotic expression for $\tau=(-N_X)^{1/2}$ in \eqref{eqn:TauExpansion}, giving an estimate for $$\A(\tbar)= C_1^{-2/5}C_2^{3/5}A\approx C_1^{-2/5}C_2^{3/5} \bigl[(-N_X)^{1/2}-\tau_\fold\bigr]|\mudot|^{-2/5},$$ which can be determined from the numerically determined $N_X(T)$. The resulting evolution of $\A(\tbar)$ is plotted for different values of $\Lambda$ in fig.~\ref{fig:TrajectoriesFullNums}. In each case, we again see that as the rate of ramping, $\dot{d}$, and hence $|\mudot|$, decreases, the trajectories of the amplitude $\A$ converge on that predicted by the asymptotic model. As expected, we also see that the numerical and asymptotic predictions diverge as $\A$ diverges: the asymptotic analysis holds only while the system remains in the vicinity of the saddle-node bifurcation (specifically, when $|\A| \ll |\mudot|^{-2/5}$). Nevertheless, as  already seen in the rescaled snap-through times plotted in fig.~\ref{fig:snaptimesvsLambda}, this period of the motion provides a bottleneck and so the resulting prediction of the duration of snap-through remain in good agreement with the numerical simulations.

\section{Summary and discussion}
\label{sec:SummaryDiscussion}
In this paper, we have presented an asymptotic reduction of the dynamic elastica equations pertinent to the snap-through of an elastic arch with time-dependent end-shortening. Our analysis allows us to understand the duration of elastic snap-through, and to show the effect of the \emph{rate} at which the end-shortening evolves. In particular, we provide semi-analytical results for the duration of snap-through, as well as the evolution of the compressive force, $-N_X$, in the vicinity of the snap-through bifurcation.

Crucially, we have shown that the point at which a rapid snap-through event occurs is delayed by the dynamic evolution of the end-shortening constraint. We have presented numerical results that show this delay and  explained the delay's origin via an asymptotically-justified reduced model, leading to asymptotic results for the size of delay that is observed in the limit of a slow, linear ramp in the control parameter ($|\mudot|\ll1$). Our results for the magnitude of the delay to the bifurcation point, expressed both in terms of the time lag, $\Delta T$, and the delay in the bifurcation parameter, $|\dMuEff|=|\mudot|\Delta T$, are summarized for the overdamped and underdamped limits in table \ref{table:Summary}. (A key feature of our asymptotic approach is that we are able to calculate the various prefactors in these scalings semi-analytically in the limit $|\mudot|\to0$. However, in a practical setting knowledge of these scalings may be sufficient.) We also determined the behaviour for intermediate damping, which depends on the size of the dimensionless damping $\nu$ in comparison to $|\mudot^{1/5}|$; see fig.~\ref{fig:snaptimesvsLambda}.

The time lag $\Delta T$ (or delay in bifurcation parameter $|\dMuEff|=|\mudot|\,\Delta T$) that we observe is the result of a delayed bifurcation \cite[][]{Su2001}; to our knowledge this is the first example of such a delayed bifurcation being discussed explicitly in the dynamics of an elastic structure. However, as we now discuss, some of the features of this delay appear to have been observed previously in related structural dynamics problems.

\subsection{Previous examples of delay in snap-through}

A possible example of  delayed bifurcation in elastic snap-through is present in the numerical results of \cite{cazzolli2019snapping}, who studied the snap through of an arch with a constant end-shortening but with one end of the arch rotated at a constant rate $\Omega$. (This problem is closely related to that studied here but with $d$ fixed and the angle of inclination $\alpha$ increasing at a constant rate $\dot{\alpha}=\Omega$.) Their numerical results show that for sufficiently small $\Omega$, snap-through occurs with a value of the end inclination angle close to that predicted by a quasi-static bifurcation analysis. However, for larger values of $\Omega$, a noticeable shift in this snapping angle is observed compared to the quasi-static bifurcation analysis \cite[see  fig.~15 of][for example]{cazzolli2019snapping}. While this observation is qualitatively consistent with the delayed bifurcation reported here, we are not able to test this conclusion quantitatively with the limited numerical data presented by \cite{cazzolli2019snapping}. 

Another example of elastic snap-through that has been studied in some detail experimentally and theoretically is that induced by the illumination of a strip of liquid crystal elastomer that is doped with azobenzene photochromes \cite[][]{Shankar2013,Korner2020}. If initially buckled as a natural arch (i.e.~with $\alpha=0$ and $d>0$ in our notation), the arch can be forced to snap to an inverted, symmetrical arch by being illuminated at, or close to, its centre: after a sufficient period of illumination, the centre of the arch is deformed sufficiently to reach an unstable configuration and snaps through \cite[as has also been achieved by the application of a point force by][]{pandey2014}. The theoretical study by \cite{Korner2020} showed that the illumination time required to induce the snap-through of the arch is approximately inversely proportional to the intensity of illumination. This inverse scaling is observed particularly at high illuminations, but at low illuminations there is  an additional delay that appears to be qualitatively consistent with the delayed bifurcation phenomenon discussed in this paper \cite[see fig.~5D of][in particular]{Korner2020}. However, this snap-through is likely to occur from a pitchfork, rather than a saddle-node, bifurcation \cite[][]{pandey2014, gomez2017critical} and so we now turn to the effect of bifurcation type on the delay that should be expected. 

\subsection{Other examples of delayed bifurcation in elastic problems}

The phenomenon of delayed bifurcation is not unique to saddle-node bifurcations and so is likely to be present in other bifurcations in solid mechanics. The most obvious candidate for these are the various forms of buckling, including dynamic Euler buckling and the dynamic torsional buckling of a rod \cite[see][for example]{Goriely2000,Zhao2019}. For completeness, we include a brief discussion of how the phenomenon of delayed bifurcation should be expected to manifest in these scenarios, though we limit our discussion to a scaling-type analysis for brevity.

\subsubsection{Pitchfork bifurcation\label{sec:Pitchfork}}

Perhaps the most common form of bifurcation in slender elastic structures is the supercritical pitchfork bifurcation seen in, for example, the Euler buckling of an inextensible rod \cite[][]{howell}. The expected normal form for a dynamic pitchfork bifurcation with constant ramping rate $\mudot$ takes the form
\beq
\frac{\upd^2A}{\upd T^2}+\nu \frac{\upd A}{\upd T}=A\left(|\mudot| T+A^2\right).
\label{eqn:PitchforkNormalForm}
\eeq   A scaling analysis of this equation can readily be performed: for the ramping and nonlinear terms to interact requires $A^2\sim |\mudot|T$, where for this discussion we use $\sim$ to mean `scales as'. In the underdamped limit, we then have that the dominant balance in \eqref{eqn:PitchforkNormalForm} must be $A/T^2\sim A^3$ so that $A\sim T^{-1}$ and hence the time lag associated with bifurcation $\Delta T\sim |\mudot|^{-1/3}$, while the associated shift in the threshold for bifurcation is $|\dMuEff|\sim |\mudot|^{2/3}$. The overdamped case can be considered through a similar scaling analysis; this shows that $\Delta T\sim (\nu/|\mudot|)^{1/2}$ and  $|\dMuEff|\sim (\nu|\mudot|)^{1/2}$. These scaling results are summarized in table \ref{table:Summary}. 

Similar scalings to those presented for the underdamped pitchfork were found in the case of dynamic Euler buckling by \cite{Kuzkin2016}; we are not aware of work in this area that considers the effect of damping, nor any that provides the overdamped scalings in this case.

\begin{table}[ht!]
\centering
\resizebox{\textwidth}{!}{%
\begin{tabular}{c | c | c | c}
Type of bifurcation & Damping? & Time delay ($\Delta T$)  & Parameter delay ($|\dMuEff|$) \\
\hline\hline & & &  \\
Saddle-node & Underdamped & $\propto |\mudot|^{-1/5}$ & $\propto |\mudot|^{4/5}$\\
Saddle-node & Overdamped & $\propto (\nu^2/|\mudot|)^{1/3}$ & $\propto (\nu |\mudot|)^{2/3}$\\
Pitchfork \& Transcritical & Underdamped & $\propto|\mudot|^{-1/3}$ & $\propto|\mudot|^{2/3}$ \\
Pitchfork \& Transcritical & Overdamped & $\propto(\nu/|\mudot|)^{1/2}$ & $\propto(\nu|\mudot|)^{1/2} $  \\
\hline\hline
\end{tabular}}
\caption{Summary of the expected delays expected in common elastic bifurcations produced by a constant ramping rate $\mudot$. In this paper we have studied in detail snap-through instability, which is of the saddle-node type with the form of amplitude equation \eqref{EigenEqn:Alpha}. A similar analysis for pitchfork bifurcations (as in buckling instability) will lead to an amplitude equation of the form \eqref{eqn:PitchforkNormalForm} while an analysis for a transcritical bifurcation will lead to \eqref{eqn:TranscriticalNormalForm}; in each case the scalings given here are expected.}
\label{table:Summary}
\end{table}

A delay in the bifurcation of a twisted elastic rod was reported by \cite{Zhao2019}. They found that the amount of the delay (i.e.~the shift in the value of the parameter at the bifurcation point) is sub-linear in $|\mudot|$, which is consistent with our scaling prediction \cite[and that of][]{Kuzkin2016}; again, more detail is required to test this similarity quantitatively.

\subsubsection{Transcritical bifurcation}

The third type of bifurcation that is seen in problems with a single real bifurcation parameter (codimension-1 bifurcations) is the transcritical bifurcation \cite[see][for example]{strogatz}. This is less common in structural stability  problems, but has been observed in the buckling of a `holey column' \cite[][]{PihlerPuzovic2016} and the elastic Rayleigh--Taylor problem \cite[][]{Chakrabarti2018}.

The normal form for a transcritical bifurcation is \cite[][]{strogatz}
\beq
\frac{\upd^2A}{\upd T^2}+\nu \frac{\upd A}{\upd T}=A\left(|\mudot| T-A \right).
\label{eqn:TranscriticalNormalForm}
\eeq  
A similar scaling analysis to that of \S\ref{sec:Pitchfork} shows that the same scalings for the size of the delay are recovered in both the underdamped and overdamped limits. This is because the interaction between the ramped parameter and the amplitude is the same in each case, with only the amplitude nonlinearity differing in the two cases (compare the first term on the right-hand side of \eqref{eqn:PitchforkNormalForm} with the corresponding term of \eqref{eqn:TranscriticalNormalForm}).

\subsection{Conclusion}

The likelihood that delayed bifurcations are ubiquitous in dynamic elastic problems may lead to both pitfalls and opportunities. As an example of a potential pitfall, we note that dynamic numerical simulations and experiments will only approximately follow the underlying static bifurcation diagram as the relevant bifurcation parameter is varied. This may be especially relevant to problems with a complicated bifurcation structure, such as the fluid-mechanical sewing machine problem \cite[][]{ChiuWebster2006,Audoly2013} and its elastic analogue \cite[][]{Jawed2014,Jawed2015}, which have multiple folds. This has been noted already in simulations of the fluid-mechanical sewing machine in which the fall height is varied slowly:   \cite{Audoly2013} observe that their numerical simulations follow the quasi-statically determined bifurcation diagram apart from a small delay at fold points caused by the small, but finite, rate of change of the bifurcation parameter \cite[see, in particular,  fig.~7a and fig.~9 of][]{Audoly2013}.

As an example of an opportunity, we note that the delay in bifurcation described in this paper may allow for detailed control of the state of bistable systems through the use of more complex time-dependent loadings than the linear ramp considered here. In particular, a non-monotonic (or slow pulse) load on a bistable structure may be used to switch between the two stable states, for example by decreasing and then increasing $d$ in the system considered here. Determining whether a slow pulse is slow and large enough to allow snap-through to occur may be amenable to an asymptotic analysis similar to that presented here, distinct from the standard energy methods \cite[][]{Simitses1990}. This approach would be particularly valuable if it allowed for a spatially uniform, but temporally varying signal to address individual `bits' in an elastic structure with multiple bistable elements \cite[][]{Seffen2006,Chung2018}. Such a development could simplify the actuation of, among other things, soft robots.

\begin{appendix}
\setcounter{figure}{0}
\setcounter{equation}{0}
\setcounter{table}{0}
\renewcommand{\thefigure}{A.\arabic{figure}}
\renewcommand{\theequation}{A.\arabic{equation}}

\section{Details of the numerical scheme \label{sec:AppendixNums}}

To solve the dynamic elastica equations, we discretize the interval $[0,1]$ using a uniform mesh with spacing $\Delta S = 1/N$ (with a fixed integer $N \geq 2$). We write $S_i = i\Delta S$ ($i = 0,1,2,\ldots,N$) for the $i$th grid point, with corresponding position vector $\mathbf{r}_i = (X_i,Y_i)^{\mathrm{T}}$. We also let $\theta_i(T)$ ($i=0,1,2,\ldots,N-1$) be the numerical approximation to $\theta(S,T)$ over the interval $(S_{i},S_i+1)$. (It should be noted that both $\theta_0$ and $\theta_N$ are imposed through boundary conditions.) We use the angles $\theta_i$ as generalised coordinates here, i.e.~we write the discretized system of equations in terms of $\theta_i$ rather than the position vectors $\mathbf{r}_i$; this approach avoids having to introduce $N$ constraints to enforce inextensibility of the centreline on each interval $(S_i,S_{i+1})$.

To obtain explicit equations for each $\theta_i$, we first integrate \eqref{eqn:HFBalND}--\eqref{eqn:VFBalND} to write the force components as
\begin{eqnarray}
N_X(S,T) & = & P(T) + \int_0^S {\pdd{X}{T} +\nu \pd{X}{T}} \upd\xi, \label{CH4eqn:integratedNX} \\ 
N_Y(S,T) & = & Q(T) + \int_0^S \pdd{Y}{T} +\nu \pd{Y}{T} \upd\xi, \label{CH4eqn:integratedNY}
\end{eqnarray}
where $P(T)$ and $Q(T)$ are unknowns that  act as Lagrange multipliers to enforce the end-shortening constraints. To achieve second-order accuracy in the convergence of our numerical scheme, we use the trapezium rule for quadrature. For a general function $f(S,T)$, we then have that for each $i = 1,2,\ldots,N-1$,
\begin{eqnarray*}
\int_0^{S_i}f(S,T)\:\upd S & \approx & \frac{\Delta S}{2}\sum_{k = 0}^{i - 1}\left[f(S_{k+1},T)+f(S_k,T)\right], \\
& = & \frac{\Delta S}{2}\left[f(0,T) +\sum_{k = 1}^{N - 1}U_{ik}f(S_k,T)\right],
\end{eqnarray*}
where in the last equality we have introduced the $(N-1)\times(N-1)$ upper triangular matrix $U = (U_{ij})$ defined by
\beq \nonumber
U_{ij} = \begin{cases}  0  & i < j,  \\  1 & i = j, \\ 2 & i > j. \end{cases}
\eeq
Setting $f(S,T) = \cos\theta(S,T)$ and $f(S,T) = \sin\theta(S,T)$ in turn, and using the fact that $\theta_0 = \alpha$, we obtain
\begin{eqnarray*}
X(S_i,T) & \approx & \frac{\Delta S}{2}\left[\cos\alpha + \sum_{k = 1}^{N - 1}U_{ik}\cos\theta_k\right], \\
Y(S_i,T) & \approx & \frac{\Delta S}{2}\left[\sin\alpha + \sum_{k = 1}^{N - 1}U_{ik}\sin\theta_k\right], \quad i = 1,2,\ldots,N-1.
\end{eqnarray*} 
It follows that 
\begin{eqnarray*}
\left(\pdd{X}{T} +\nu \pd{X}{T} \right)\bigg\rvert_{S = S_i} & \approx & -\frac{\Delta S}{2}\sum_{k = 1}^{N-1}U_{ik}\left[\left(\ddd{\theta_k}{T} +\nu \dd{\theta_k}{T} \right)\sin\theta_k+\left(\dd{\theta_k}{T}\right)^2\cos\theta_k\right], \\
\left(\pdd{Y}{T} +\nu \pd{Y}{T} \right)\bigg\rvert_{S = S_i} & \approx & \frac{\Delta S}{2}\sum_{k = 1}^{N-1}U_{ik}\left[\left(\ddd{\theta_k}{T} +\nu \dd{\theta_k}{T} \right)\cos\theta_k-\left(\dd{\theta_k}{T}\right)^2\sin\theta_k\right].
\end{eqnarray*}
Substituting the above into \eqref{CH4eqn:integratedNX}, we may then approximate the force component $N_X$ as (making use of $X(0,T) = 0$)
\begin{eqnarray}
N_X(S_i,T)& \approx & P(T) +\frac{\Delta S}{2}\sum_{j = 1}^{N-1}U_{ij}\left(\pdd{X}{T} +\nu \pd{X}{T} \right)\bigg\rvert_{S = S_j}, \nonumber \\
& \approx & P(T)-\frac{\Delta S^2}{4}\sum_{j = 1}^{N-1}U_{ij}\sum_{k = 1}^{N-1}U_{jk}\left[\left(\ddd{\theta_k}{T} +\nu \dd{\theta_k}{T} \right)\sin\theta_k+\left(\dd{\theta_k}{T}\right)^2\cos\theta_k\right]. \nonumber \\ \label{CH4eqn:NXscheme}
\end{eqnarray}
Similarly, using \eqref{CH4eqn:integratedNY} and $Y(0,T) = 0$, we have
\beq
N_Y(S_i,T) \approx Q(T) + \frac{\Delta S^2}{4}\sum_{j = 1}^{N-1}U_{ij}\sum_{k = 1}^{N-1}U_{jk}\left[\left(\ddd{\theta_k}{T} +\nu \dd{\theta_k}{T} \right)\cos\theta_k-\left(\dd{\theta_k}{T}\right)^2\sin\theta_k\right].  \label{CH4eqn:NYscheme}
\eeq

We approximate the $\partial^2\theta/\partial S^2$ term appearing in the moment balance \eqref{eqn:ElasticaND} using a second-order centered difference on the interior grid points. Upon substituting the approximations \eqref{CH4eqn:NXscheme}--\eqref{CH4eqn:NYscheme}, and making use of the addition formulae for $\sin(\theta_i-\theta_k)$ and $\cos(\theta_i-\theta_k)$, equation \eqref{eqn:ElasticaND} becomes
\begin{eqnarray}
&& \frac{\theta_{i+1}-2\theta_{i}+\theta_{i-1}}{\Delta S^2} - P\sin\theta_i+ Q\cos\theta_i  \nonumber \\
&& = - \frac{\Delta S^2}{4}\sum_{j = 1}^{N-1}U_{ij}\sum_{k = 1}^{N-1}U_{jk}\left[\left(\ddd{\theta_k}{T} +\nu \dd{\theta_k}{T} \right)\cos(\theta_i-\theta_k)+\left(\dd{\theta_k}{T}\right)^2\sin(\theta_i-\theta_k)\right],  \nonumber \\
\label{CH4eqn:thetaequations}
\end{eqnarray}for $i = 1,2,\ldots,N-1$. The boundary conditions \eqref{eqn:ThetaBC} give
\beq
\theta_0 = \alpha,\quad \theta_N = 0. \label{CH4eqn:thetaBCS}
\eeq
The imposed end-shortening $X(1,T) = 1-d$ and $Y(1,T) = 0$ (recall \eqref{eqn:XYBC1}-\eqref{eqn:XYBC2}) leads to the algebraic constraints
\begin{eqnarray}
&& \frac{\Delta S}{2}\left[\cos\alpha + 1 + 2\sum_{k = 1}^{N - 1}\cos\theta_k\right] = 1-\left[d_c(\alpha)+\dot{d}\,T\right], \nonumber \\
&& \frac{\Delta S}{2}\left[\sin\alpha + 2\sum_{k = 1}^{N - 1}\sin\theta_k\right] = 0. \label{CH4eqn:schemeconstraints}
\end{eqnarray}

To write the system in matrix form, we introduce the column vectors of length $(N-1)$ 
\begin{eqnarray*}
\mathbf{\Theta} &= & (\theta_1,\theta_2,\ldots,\theta_{N-1})^{\mathrm{T}}, \\
\mathbf{s} &= & (\sin\theta_1,\sin\theta_2,\ldots,\sin\theta_{N-1})^{\mathrm{T}}, \\
\mathbf{c} &= & (\cos\theta_1,\cos\theta_2,\ldots,\cos\theta_{N-1})^{\mathrm{T}}, \\
\mathbf{e}_1 &= & (1,0,\ldots,0)^{\mathrm{T}}.
\end{eqnarray*}
We define the $(N-1)\times(N-1)$ upper triangular matrices $A = (A_{ij})$ and $B = (B_{ij})$ where 
\begin{eqnarray*}
A_{ij} &= & \cos\left(\theta_i-\theta_j\right)(U^2)_{ij}, \\
B_{ij} &= & \sin\left(\theta_i-\theta_j\right)(U^2)_{ij}.
\end{eqnarray*}
We also define the $(N-1)\times (N-1)$  tridiagonal matrix 
\beq \nonumber
D = \frac{1}{\Delta S^2}\begin{pmatrix}
-2 & 1 & \\
1 & -2 & 1 \\
& \ddots & \ddots & \ddots \\
&& 1 & -2 & 1 \\
& & & 1 & -2  \\
\end{pmatrix} .
\eeq
The system of equations \eqref{CH4eqn:thetaequations} can then be written as 
\beq
A\ddd{\mathbf{\Theta}}{T} + \left[\nu A + B\:\mathrm{diag}\left(\dd{\mathbf{\Theta}}{T}\right)\right]\dd{\mathbf{\Theta}}{T} = \frac{4}{\Delta S^2}\left[P\mathbf{s}-Q\mathbf{c}-\frac{\alpha}{\Delta S^2}\mathbf{e}_1 - D\mathbf{\Theta}\right], \label{CH4eqn:matrixform}
\eeq
where we write $\mathrm{diag}(\mathbf{a})$ for the diagonal matrix whose diagonal entries are the entries of the vector $\mathbf{a}$.  Together with the constraints \eqref{CH4eqn:schemeconstraints}, these equations constitute a system of differential algebraic equations (DAEs) since the unknown functions $P(T)$ and $Q(T)$ do not explicitly enter  \eqref{CH4eqn:schemeconstraints}. It is convenient to reduce the index of the system by differentiating \eqref{CH4eqn:schemeconstraints} once, which can then be written in the form
\begin{eqnarray}
&& \Delta S\:\mathbf{s}^{\mathrm{T}}\dd{\mathbf{\Theta}}{T} = \dot{d}, \label{eqn:ddot1}\\
&& \Delta S\:\mathbf{c}^{\mathrm{T}}\dd{\mathbf{\Theta}}{T} = 0\label{eqn:ddot2}.
\end{eqnarray}
The numerical solution of the system \eqref{CH4eqn:matrixform}--\eqref{eqn:ddot2} then satisfies \eqref{CH4eqn:schemeconstraints} (to within numerical tolerances) provided that the initial data are consistent with  \eqref{CH4eqn:schemeconstraints}. (Since the arch starts from rest but we implement the end-shortening constraint through \eqref{eqn:ddot1}, we introduce a time-varying $\dot{d}$ that accelerates rapidly from 0 to the prescribed constant value $\dot{d}_0$, namely $$\dot{d} = \dot{d}_0 \cdot (1-e^{-T/T_{\mathrm{decay}}}),$$ where $T_{\mathrm{decay}} \ll T_\start$.) Together with \eqref{CH4eqn:matrixform}, the DAE system can then easily be written in mass-matrix form (with singular mass matrix) and integrated numerically using  \textsc{matlab}'s ODE solvers. We note that it is possible to obtain explicit expressions for the Lagrange multipliers  $P$ and $Q$ (up to the solution of a linear system) and hence avoid a singular mass matrix \citep{ruhoff1996efficient}; alternatively, using the impetus-striction method, the system can be rephrased as an unconstrained Hamiltonian system in which the constraints \eqref{CH4eqn:schemeconstraints} are automatically satisfied \citep{dichmann1996impetus}. However, we found that the formulation here can be efficiently integrated using the routine \texttt{ode15s}, once the sparsity pattern of the mass matrix is specified (for example the matrix $A$ is upper triangular); simulations using $N= 100$ typically complete in under a minute on a laptop computer. We also note that the matrices $A$ and $B$ can be efficiently constructed by applying the routine \texttt{meshgrid} to the vector $\mathbf{\Theta}$. 

\subsection{Equilibrium solutions}
Setting the time derivatives to zero in \eqref{CH4eqn:matrixform} shows that equilibrium solutions satisfy 
\beq
D\mathbf{\Theta} = P\mathbf{s}-Q\mathbf{c}-\frac{\alpha}{\Delta S^2}\mathbf{e}_1, \label{CH4eqn:dteadymatrixform} 
\eeq
together with \eqref{CH4eqn:schemeconstraints}. We solve these algebraic equations in \textsc{matlab} using the routine \texttt{fsolve} to determine $\mathbf{\Theta}$ and the unknown force components $P$ and $Q$. To obtain a response diagram as we vary the dimensionless end-shortening $d=\Delta L/L$, we implement the numerical solver together with a simple continuation algorithm. Rather than controlling $d$, however, we instead control $P$ and solve for the corresponding value of $d$ at each stage. This allows us to use a simple continuation algorithm that continues past the saddle-node bifurcation without any convergence issues. (Controlling instead $d$ produces  a vertical fold in the bifurcation diagram; near this fold a small change in $d$ produces a large change in the solution.) In dealing with the different solution branches, we use the analytical solutions predicted by beam theory (derived in \S\ref{sec:QuasiLinStatics}) as initial guesses, relating these to $\mathbf{\Theta}$, $P$ and $Q$. 

\subsection{Convergence tests}
\begin{figure}
\centering
\includegraphics[width=1.0\textwidth]{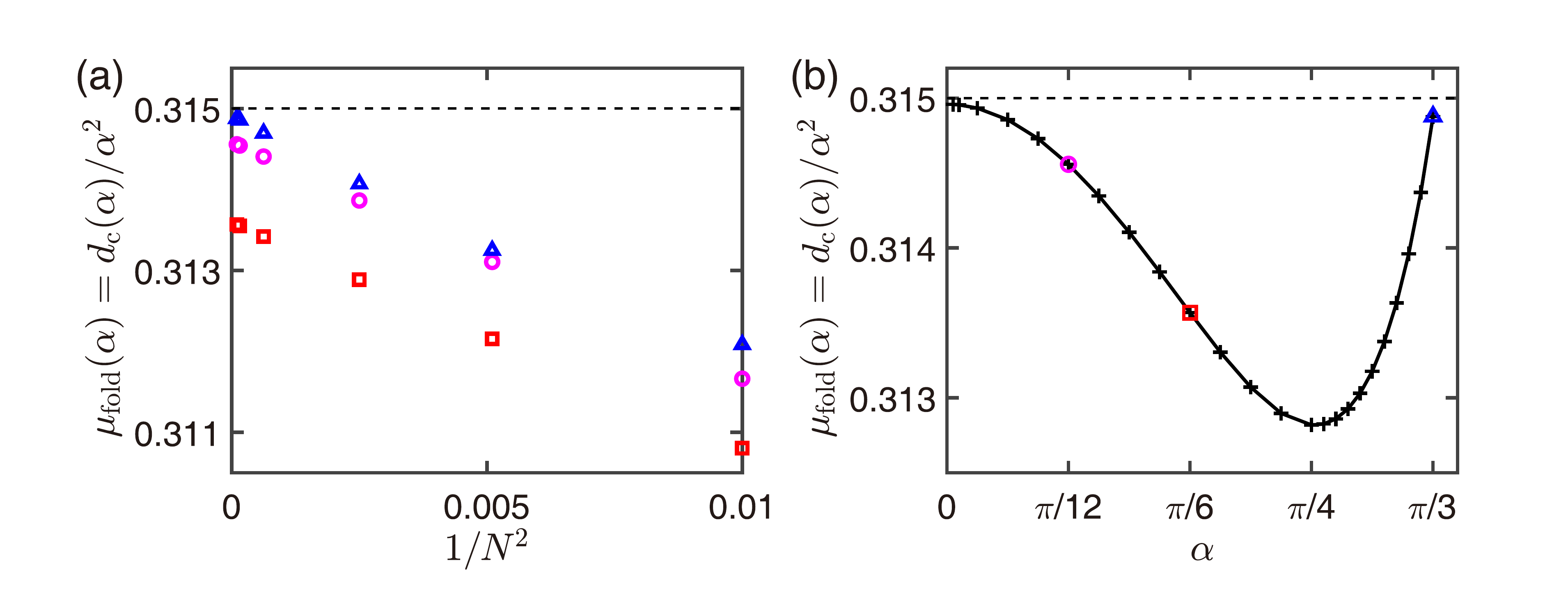}
\caption{Results of a convergence analysis for the static problem. (a) The values of $\mu_\fold(\alpha)$ determined numerically from the static code depend quadratically on the grid spacing $\Delta S=1/N$ for $\alpha=\pi/12$ (circles), $\alpha=\pi/6$ (squares) and $\alpha=\pi/3$ (triangles). (This quadratic convergence is consistent with our use of a second-order finite-difference method.) In each case, the converged value of $\mu_\fold$ for each $\alpha$ is very close to the value predicted by our quasi-linear theory, \eqref{eqn:FoldProperties}, namely $\mu_\fold\approx0.3150$ (dashed line). (b) The value of $\mu_\fold(\alpha)$ determined with $N=100$ grid points  shows that the numerically determined value lies within $0.7\%$ of the analytical value $\mu_\fold\approx0.315$ (determined from the quasi-linear theory), for  $\alpha\lesssim\pi/3$.
\label{fig:ConvergenceTest}}
\end{figure}

The numerical solutions of the equilibrium problem may be used to check the convergence of our numerical scheme with increasing numbers of grid points, $N$. In particular, we calculate the critical value of the end-shortening, $d_\fold$, for different arch inclinations $\alpha$ and discretization levels $N$. From this, we can readily compute a numerical estimate of the value of $\mu_\fold=d_\fold/\alpha^2$ with different values of $\alpha$ and $N$. Figure \ref{fig:ConvergenceTest}a shows that as $N$ increases, the value of $\mu_\fold$ converges; in particular, we see that this convergence is second order in the grid-spacing $\Delta S=1/N$, as should be expected because we use second order finite differences throughout. The plot in fig.~\ref{fig:ConvergenceTest}b shows that there is a weak dependence of the converged ($N\to\infty$) value of $\mu_\fold$ on the value of $\alpha$. This is to be expected because the analytical value of $\mu_\fold\approx 0.315$, given in \eqref{eqn:FoldProperties}, is only valid in the limit of small deflections, $\alpha\ll1$. Figure \ref{fig:ConvergenceTest}b confirms that the analytical result \eqref{eqn:FoldProperties} is recovered as $\alpha\to0$ and, further, that, for $\alpha\in (0,\pi/3]$, the true value lies within $0.7\%$ of the analytical value of \eqref{eqn:FoldProperties}.

\end{appendix}

\section*{Acknowledgments}

The research leading to these results has received funding from the Royal Society through a Newton International Fellowship (ML), Peterhouse, Cambridge (MG), the European Research Council under the European Union's Horizon 2020 Programme / ERC Grant Agreement no.~637334 (DV) and the Leverhulme Trust through a Philip Leverhulme Prize (DV). We are grateful to Benny Davidovitch for introducing us to the terminology of delayed bifurcation.

\section*{References}


\begin{thebibliography}{53}
\expandafter\ifx\csname natexlab\endcsname\relax\def\natexlab#1{#1}\fi
\providecommand{\url}[1]{\texttt{#1}}
\providecommand{\href}[2]{#2}
\providecommand{\path}[1]{#1}
\providecommand{\DOIprefix}{doi:}
\providecommand{\ArXivprefix}{arXiv:}
\providecommand{\URLprefix}{URL: }
\providecommand{\Pubmedprefix}{pmid:}
\providecommand{\doi}[1]{\href{http://dx.doi.org/#1}{\path{#1}}}
\providecommand{\Pubmed}[1]{\href{pmid:#1}{\path{#1}}}
\providecommand{\bibinfo}[2]{#2}
\ifx\xfnm\relax \def\xfnm[#1]{\unskip,\space#1}\fi
\bibitem[{Forterre et~al.(2005)Forterre, Skotheim, Dumais, and
  Mahadevan}]{forterre2005}
\bibinfo{author}{Y.~Forterre}, \bibinfo{author}{J.~M. Skotheim},
  \bibinfo{author}{J.~Dumais}, \bibinfo{author}{L.~Mahadevan},
\newblock \bibinfo{title}{How the {V}enus flytrap snaps},
\newblock \bibinfo{journal}{Nature} \bibinfo{volume}{433}
  (\bibinfo{year}{2005}) \bibinfo{pages}{421--425}.
\bibitem[{Skotheim and Mahadevan(2005)}]{Skotheim2005}
\bibinfo{author}{J.~M. Skotheim}, \bibinfo{author}{L.~Mahadevan},
\newblock \bibinfo{title}{Physical limits and design principles for plant and
  fungal movements},
\newblock \bibinfo{journal}{Science} \bibinfo{volume}{308}
  (\bibinfo{year}{2005}) \bibinfo{pages}{1309--1310}.
\bibitem[{Smith et~al.(2011)Smith, Yanega, and Ruina}]{Smith2011}
\bibinfo{author}{M.~L. Smith}, \bibinfo{author}{G.~M. Yanega},
  \bibinfo{author}{A.~Ruina},
\newblock \bibinfo{title}{Elastic instability model of rapid beak closure in
  hummingbirds},
\newblock \bibinfo{journal}{J. Theo. Bio.} \bibinfo{volume}{282}
  (\bibinfo{year}{2011}) \bibinfo{pages}{41--51}.
\bibitem[{Hu and Burgue{\~n}o(2015)}]{hu2015buckling}
\bibinfo{author}{N.~Hu}, \bibinfo{author}{R.~Burgue{\~n}o},
\newblock \bibinfo{title}{Buckling-induced smart applications: {R}ecent
  advances and trends},
\newblock \bibinfo{journal}{Smart Mater. Struct.} \bibinfo{volume}{24}
  (\bibinfo{year}{2015}) \bibinfo{pages}{063001}.
\bibitem[{Holmes and Crosby(2007)}]{Holmes2007}
\bibinfo{author}{D.~P. Holmes}, \bibinfo{author}{A.~J. Crosby},
\newblock \bibinfo{title}{Snapping surfaces},
\newblock \bibinfo{journal}{Adv. Mater.} \bibinfo{volume}{19}
  (\bibinfo{year}{2007}) \bibinfo{pages}{3589--3593}.
\bibitem[{Gon\c{c}alves et~al.(2003)Gon\c{c}alves, Pamplona, Teixeira,
  Jerusalmi, Cestari, and Leirner}]{Goncalves2003}
\bibinfo{author}{P.~B. Gon\c{c}alves}, \bibinfo{author}{D.~Pamplona},
  \bibinfo{author}{P.~B. Teixeira}, \bibinfo{author}{R.~L. Jerusalmi},
  \bibinfo{author}{I.~A. Cestari}, \bibinfo{author}{A.~A. Leirner},
\newblock \bibinfo{title}{Dynamic non-linear behavior and stability of a
  ventricular assist device},
\newblock \bibinfo{journal}{Int. J. Solids Struct.} \bibinfo{volume}{40}
  (\bibinfo{year}{2003}) \bibinfo{pages}{5017--5035}.
\bibitem[{Rafsanjani et~al.(2015)Rafsanjani, Akbarzadeh, and
  Pasini}]{Rafsanjani2015}
\bibinfo{author}{A.~Rafsanjani}, \bibinfo{author}{A.~Akbarzadeh},
  \bibinfo{author}{D.~Pasini},
\newblock \bibinfo{title}{Snapping mechanical metamaterials under tension},
\newblock \bibinfo{journal}{Adv. Mater.} \bibinfo{volume}{27}
  (\bibinfo{year}{2015}) \bibinfo{pages}{5931--5935}.
\bibitem[{Janbaz et~al.(2020)Janbaz, Narooei, van Manen, and
  Zadpoor}]{Janbaz2020}
\bibinfo{author}{S.~Janbaz}, \bibinfo{author}{K.~Narooei},
  \bibinfo{author}{T.~van Manen}, \bibinfo{author}{A.~A. Zadpoor},
\newblock \bibinfo{title}{Strain rate–dependent mechanical metamaterials},
\newblock \bibinfo{journal}{Sci. Adv.} \bibinfo{volume}{6}
  (\bibinfo{year}{2020}) \bibinfo{pages}{eaba0616}.
\bibitem[{Santer(2010)}]{Santer2010}
\bibinfo{author}{M.~Santer},
\newblock \bibinfo{title}{Self-actuated snap back of viscoelastic pulsing
  structures},
\newblock \bibinfo{journal}{Int. J. Solids Struct.} \bibinfo{volume}{47}
  (\bibinfo{year}{2010}) \bibinfo{pages}{3263--3271}.
\bibitem[{Brinkmeyer et~al.(2012)Brinkmeyer, Santer, Pirrera, and
  Weaver}]{brinkmeyer2012}
\bibinfo{author}{A.~Brinkmeyer}, \bibinfo{author}{M.~Santer},
  \bibinfo{author}{A.~Pirrera}, \bibinfo{author}{P.~M. Weaver},
\newblock \bibinfo{title}{Pseudo-bistable self-actuated domes for morphing
  applications},
\newblock \bibinfo{journal}{Int. J. Solids Struct.} \bibinfo{volume}{49}
  (\bibinfo{year}{2012}) \bibinfo{pages}{1077--1087}.
\bibitem[{Brinkmeyer et~al.(2013)Brinkmeyer, Pirrera, Santer, and
  Weaver}]{brinkmeyer2013}
\bibinfo{author}{A.~Brinkmeyer}, \bibinfo{author}{A.~Pirrera},
  \bibinfo{author}{M.~Santer}, \bibinfo{author}{P.~M. Weaver},
\newblock \bibinfo{title}{Pseudo-bistable pre-stressed morphing composite
  panels},
\newblock \bibinfo{journal}{Int. J. Solids Struct.} \bibinfo{volume}{50}
  (\bibinfo{year}{2013}) \bibinfo{pages}{1033--1043}.
\bibitem[{Gomez et~al.(2019)Gomez, Moulton, and Vella}]{Gomez2019}
\bibinfo{author}{M.~Gomez}, \bibinfo{author}{D.~E. Moulton},
  \bibinfo{author}{D.~Vella},
\newblock \bibinfo{title}{Dynamics of viscoelastic snap-through},
\newblock \bibinfo{journal}{J. Mech. Phys. Solids} \bibinfo{volume}{124}
  (\bibinfo{year}{2019}) \bibinfo{pages}{781--813}.
\bibitem[{Urbach and Efrati(2020)}]{Urbach2020}
\bibinfo{author}{E.~Y. Urbach}, \bibinfo{author}{E.~Efrati},
\newblock \bibinfo{title}{Predicting delayed instabilities in viscoelastic
  solids},
\newblock \bibinfo{journal}{Sci. Adv.} \bibinfo{volume}{6}
  (\bibinfo{year}{2020}) \bibinfo{pages}{eabb2948}.
\bibitem[{Pandey et~al.(2014)Pandey, Moulton, Vella, and Holmes}]{pandey2014}
\bibinfo{author}{A.~Pandey}, \bibinfo{author}{D.~E. Moulton},
  \bibinfo{author}{D.~Vella}, \bibinfo{author}{D.~P. Holmes},
\newblock \bibinfo{title}{Dynamics of snapping beams and jumping poppers},
\newblock \bibinfo{journal}{Europhys. Lett.} \bibinfo{volume}{105}
  (\bibinfo{year}{2014}) \bibinfo{pages}{24001}.
\bibitem[{Gomez et~al.(2017)Gomez, Moulton, and Vella}]{gomez2017critical}
\bibinfo{author}{M.~Gomez}, \bibinfo{author}{D.~E. Moulton},
  \bibinfo{author}{D.~Vella},
\newblock \bibinfo{title}{Critical slowing down in purely elastic
  ‘snap-through’ instabilities},
\newblock \bibinfo{journal}{Nat. Phys.} \bibinfo{volume}{13}
  (\bibinfo{year}{2017}) \bibinfo{pages}{142--145}.
\bibitem[{Sano and Wada(2018)}]{Sano2018}
\bibinfo{author}{T.~Sano}, \bibinfo{author}{H.~Wada},
\newblock \bibinfo{title}{Snap-buckling in asymmetrically constrained elastic
  strips},
\newblock \bibinfo{journal}{Phys. Rev. E} \bibinfo{volume}{97}
  (\bibinfo{year}{2018}) \bibinfo{pages}{013002}.
\bibitem[{Bushnell(1981)}]{bushnell1981}
\bibinfo{author}{D.~Bushnell},
\newblock \bibinfo{title}{Buckling of shells---pitfall for designers},
\newblock \bibinfo{journal}{AIAA J.} \bibinfo{volume}{19}
  (\bibinfo{year}{1981}) \bibinfo{pages}{1183--1226}.
\bibitem[{Strogatz(2014)}]{strogatz}
\bibinfo{author}{S.~H. Strogatz}, \bibinfo{title}{Nonlinear Dynamics and
  Chaos}, \bibinfo{publisher}{Westview Press, Boulder, CO},
  \bibinfo{year}{2014}.
\bibitem[{Gomez(2018)}]{gomezthesis}
\bibinfo{author}{M.~Gomez}, \bibinfo{title}{Ghosts and bottlenecks in elastic
  snap-through}, Ph.D. thesis, University of Oxford, \bibinfo{year}{2018}.
\bibitem[{Epstein et~al.(2015)Epstein, Yoon, Madhukar, Hsia, and
  Braun}]{epstein2015}
\bibinfo{author}{E.~Epstein}, \bibinfo{author}{J.~Yoon},
  \bibinfo{author}{A.~Madhukar}, \bibinfo{author}{K.~J. Hsia},
  \bibinfo{author}{P.~V. Braun},
\newblock \bibinfo{title}{Colloidal particles that rapidly change shape via
  elastic instabilities},
\newblock \bibinfo{journal}{Small} \bibinfo{volume}{11} (\bibinfo{year}{2015})
  \bibinfo{pages}{6051--6057}.
\bibitem[{Lee et~al.(2010)Lee, Xia, and Fang}]{lee2010}
\bibinfo{author}{H.~Lee}, \bibinfo{author}{C.~Xia}, \bibinfo{author}{N.~X.
  Fang},
\newblock \bibinfo{title}{First jump of a mircogel; actuation speed enhancement
  by elastic instability},
\newblock \bibinfo{journal}{Soft Matter} \bibinfo{volume}{6}
  (\bibinfo{year}{2010}) \bibinfo{pages}{4342--4345}.
\bibitem[{Loukaides et~al.(2014)Loukaides, Smoukov, and Seffen}]{Loukaides2014}
\bibinfo{author}{E.~G. Loukaides}, \bibinfo{author}{S.~K. Smoukov},
  \bibinfo{author}{K.~A. Seffen},
\newblock \bibinfo{title}{Magnetic actuation and transition shapes of a
  bistable spherical cap},
\newblock \bibinfo{journal}{Int. J. Smart Nano Mater.} \bibinfo{volume}{5}
  (\bibinfo{year}{2014}) \bibinfo{pages}{270--282}.
\bibitem[{Seffen and Vidoli(2016)}]{Seffen2016}
\bibinfo{author}{K.~A. Seffen}, \bibinfo{author}{S.~Vidoli},
\newblock \bibinfo{title}{Eversion of bistable shells under magnetic actuation:
  a model of nonlinear shapes},
\newblock \bibinfo{journal}{Smart Mater. Struct.} \bibinfo{volume}{25}
  (\bibinfo{year}{2016}) \bibinfo{pages}{065010}.
\bibitem[{Gomez et~al.(2017)Gomez, Moulton, and Vella}]{gomez2017b}
\bibinfo{author}{M.~Gomez}, \bibinfo{author}{D.~E. Moulton},
  \bibinfo{author}{D.~Vella},
\newblock \bibinfo{title}{Passive control of viscous flow via elastic
  snap-through},
\newblock \bibinfo{journal}{Phys. Rev. Lett.} \bibinfo{volume}{119}
  (\bibinfo{year}{2017}) \bibinfo{pages}{144502}.
\bibitem[{Arena et~al.(2017)Arena, Groh, Brinkmeyer, Theunissen, Weaver, and
  Pirrera}]{Arena2017}
\bibinfo{author}{G.~Arena}, \bibinfo{author}{R.~M.~J. Groh},
  \bibinfo{author}{A.~Brinkmeyer}, \bibinfo{author}{R.~Theunissen},
  \bibinfo{author}{P.~Weaver}, \bibinfo{author}{A.~Pirrera},
\newblock \bibinfo{title}{Adaptive compliant structures for flow regulation},
\newblock \bibinfo{journal}{Proc. R. Soc. A} \bibinfo{volume}{473}
  (\bibinfo{year}{2017}) \bibinfo{pages}{20170334}.
\bibitem[{Tredicce et~al.(2004)Tredicce, Lippi, Mandel, Charasse, Chevalier,
  and Picqu{\'e}}]{tredicce2004}
\bibinfo{author}{J.~R. Tredicce}, \bibinfo{author}{G.~L. Lippi},
  \bibinfo{author}{P.~Mandel}, \bibinfo{author}{B.~Charasse},
  \bibinfo{author}{A.~Chevalier}, \bibinfo{author}{B.~Picqu{\'e}},
\newblock \bibinfo{title}{Critical slowing down at a bifurcation},
\newblock \bibinfo{journal}{Am. J. Phys.} \bibinfo{volume}{72}
  (\bibinfo{year}{2004}) \bibinfo{pages}{799--809}.
\bibitem[{Majumdar et~al.(2013)Majumdar, Ockendon, Howell, and
  Surovyatkina}]{majumdar2013}
\bibinfo{author}{A.~Majumdar}, \bibinfo{author}{J.~Ockendon},
  \bibinfo{author}{P.~Howell}, \bibinfo{author}{E.~Surovyatkina},
\newblock \bibinfo{title}{Transitions through critical temperatures in nematic
  liquid crystals},
\newblock \bibinfo{journal}{Phys. Rev. E} \bibinfo{volume}{88}
  (\bibinfo{year}{2013}) \bibinfo{pages}{022501}.
\bibitem[{Su(2001)}]{Su2001}
\bibinfo{author}{J.~Su},
\newblock \bibinfo{title}{The phenomenon of delayed bifurcation and its
  analyses},
\newblock in: \bibinfo{editor}{C.~K. R.~T. Jones}, \bibinfo{editor}{A.~I.
  Khibnik} (Eds.), \bibinfo{booktitle}{Multiple-Time-Scale Dynamical Systems.},
  \bibinfo{publisher}{Springer}, \bibinfo{year}{2001}, p. \bibinfo{pages}{203}.
\bibitem[{Maddocks(1987)}]{maddocks1987}
\bibinfo{author}{J.~H. Maddocks},
\newblock \bibinfo{title}{Stability and folds},
\newblock \bibinfo{journal}{Arch. Rational Mech. Anal.} \bibinfo{volume}{99}
  (\bibinfo{year}{1987}) \bibinfo{pages}{301--328}.
\bibitem[{Schiesser and Griffiths(2009)}]{Schiesser2009}
\bibinfo{author}{W.~Schiesser}, \bibinfo{author}{G.~W. Griffiths},
  \bibinfo{title}{A Compendium of Partial Differential Equation Models: Method
  of Lines Analysis with Matlab}, \bibinfo{publisher}{Cambridge University
  Press}, \bibinfo{year}{2009}.
\bibitem[{Keener(1988)}]{Keener}
\bibinfo{author}{J.~P. Keener}, \bibinfo{title}{Principles of Applied
  Mathematics}, \bibinfo{publisher}{Addison-Wesley, Boston, MA},
  \bibinfo{year}{1988}.
\bibitem[{Erneux and Laplante(1989)}]{erneux1989}
\bibinfo{author}{T.~Erneux}, \bibinfo{author}{J.~P. Laplante},
\newblock \bibinfo{title}{Jump transition due to a time-dependent bifurcation
  parameter in the bistable ioadate--arsenous acid reaction},
\newblock \bibinfo{journal}{J. Chem. Phys.} \bibinfo{volume}{90}
  (\bibinfo{year}{1989}) \bibinfo{pages}{6129--6134}.
\bibitem[{Laplante et~al.(1991)Laplante, Erneux, and Georgiou}]{laplante1991}
\bibinfo{author}{J.~P. Laplante}, \bibinfo{author}{T.~Erneux},
  \bibinfo{author}{M.~Georgiou},
\newblock \bibinfo{title}{Jump transition due to a time-dependent bifurcation
  parameter: An experimental, numerical, and analytical study of the bistable
  iodate--arsenous acid reaction},
\newblock \bibinfo{journal}{J. Chem. Phys.} \bibinfo{volume}{94}
  (\bibinfo{year}{1991}) \bibinfo{pages}{371--378}.
\bibitem[{Zhu et~al.(2015)Zhu, Kuske, and Erneux}]{zhu2015}
\bibinfo{author}{J.~Zhu}, \bibinfo{author}{R.~Kuske},
  \bibinfo{author}{T.~Erneux},
\newblock \bibinfo{title}{Tipping points near a delayed saddle node bifurcation
  with periodic forcing},
\newblock \bibinfo{journal}{SIAM J. Appl. Dyn. Syst.} \bibinfo{volume}{14}
  (\bibinfo{year}{2015}) \bibinfo{pages}{2030--2068}.
\bibitem[{Li et~al.(2019)Li, Ye, Qian, and Huang}]{li2019}
\bibinfo{author}{J.~H. Li}, \bibinfo{author}{F.~X.-F. Ye},
  \bibinfo{author}{H.~Qian}, \bibinfo{author}{S.~Huang},
\newblock \bibinfo{title}{Time-dependent saddle--node bifurcation: Breaking
  time and the point of no return in a non-autonomous model of critical
  transitions},
\newblock \bibinfo{journal}{Physica D} \bibinfo{volume}{395}
  (\bibinfo{year}{2019}) \bibinfo{pages}{7--14}.
\bibitem[{Cazzolli and Dal~Corso(2019)}]{cazzolli2019snapping}
\bibinfo{author}{A.~Cazzolli}, \bibinfo{author}{F.~Dal~Corso},
\newblock \bibinfo{title}{Snapping of elastic strips with controlled ends},
\newblock \bibinfo{journal}{Int. J. Solids Struct.} \bibinfo{volume}{162}
  (\bibinfo{year}{2019}) \bibinfo{pages}{285--303}.
\bibitem[{{Ravi Shankar} et~al.(2013){Ravi Shankar}, Smith, Tondiglia, Lee,
  {McConney}, Wang, Tan, and White}]{Shankar2013}
\bibinfo{author}{M.~{Ravi Shankar}}, \bibinfo{author}{M.~L. Smith},
  \bibinfo{author}{V.~P. Tondiglia}, \bibinfo{author}{K.~M. Lee},
  \bibinfo{author}{M.~E. {McConney}}, \bibinfo{author}{D.~H. Wang},
  \bibinfo{author}{L.-S. Tan}, \bibinfo{author}{T.~J. White},
\newblock \bibinfo{title}{Contactless, photoinitiated snap-through in
  azobenzene-functionalized polymers},
\newblock \bibinfo{journal}{Proc. Natl Acad. Sci. USA} \bibinfo{volume}{110}
  (\bibinfo{year}{2013}) \bibinfo{pages}{18792–--18797}.
\bibitem[{Korner et~al.(2020)Korner, Kuenstler, Hayward, Audoly, and
  Bhattacharya}]{Korner2020}
\bibinfo{author}{K.~Korner}, \bibinfo{author}{A.~S. Kuenstler},
  \bibinfo{author}{R.~C. Hayward}, \bibinfo{author}{B.~Audoly},
  \bibinfo{author}{K.~Bhattacharya},
\newblock \bibinfo{title}{A nonlinear beam model of photomotile structures},
\newblock \bibinfo{journal}{Proc. Natl Acad. Sci. USA} \bibinfo{volume}{117}
  (\bibinfo{year}{2020}) \bibinfo{pages}{9762–--9770}.
\bibitem[{Goriely and Tabor(2000)}]{Goriely2000}
\bibinfo{author}{A.~Goriely}, \bibinfo{author}{M.~Tabor},
\newblock \bibinfo{title}{The nonlinear dynamics of filaments},
\newblock \bibinfo{journal}{Nonlin. Dyn.} \bibinfo{volume}{21}
  (\bibinfo{year}{2000}) \bibinfo{pages}{101--133}.
\bibitem[{Zhao and {van der Heijden}(2019)}]{Zhao2019}
\bibinfo{author}{X.~W. Zhao}, \bibinfo{author}{G.~H.~M. {van der Heijden}},
\newblock \bibinfo{title}{Dynamic torsional buckling: Prebuckling waves and
  delayed instability},
\newblock \bibinfo{journal}{Commun. Nonlinear Sci. Numer. Simulat.}
  \bibinfo{volume}{69} (\bibinfo{year}{2019}) \bibinfo{pages}{360--369}.
\bibitem[{Howell et~al.(2009)Howell, Kozyreff, and Ockendon}]{howell}
\bibinfo{author}{P.~Howell}, \bibinfo{author}{G.~Kozyreff},
  \bibinfo{author}{J.~Ockendon}, \bibinfo{title}{Applied Solid Mechanics},
  \bibinfo{publisher}{Cambridge University Press, Cambridge},
  \bibinfo{year}{2009}.
\bibitem[{Kuzkin and Dannert(2016)}]{Kuzkin2016}
\bibinfo{author}{V.~A. Kuzkin}, \bibinfo{author}{M.~M. Dannert},
\newblock \bibinfo{title}{Buckling of a column under a constant speed
  compression: a dynamic correction to the euler formula},
\newblock \bibinfo{journal}{Acta Mech.} \bibinfo{volume}{227}
  (\bibinfo{year}{2016}) \bibinfo{pages}{1645--1652}.
\bibitem[{Pihler-Puzov\'{i}c et~al.(2016)Pihler-Puzov\'{i}c, Hazel, and
  Mullin}]{PihlerPuzovic2016}
\bibinfo{author}{D.~Pihler-Puzov\'{i}c}, \bibinfo{author}{A.~L. Hazel},
  \bibinfo{author}{T.~Mullin},
\newblock \bibinfo{title}{Buckling of a holey column},
\newblock \bibinfo{journal}{Soft Matter} \bibinfo{volume}{12}
  (\bibinfo{year}{2016}) \bibinfo{pages}{7112--7118}.
\bibitem[{Chakrabarti et~al.(2018)Chakrabarti, Mora, Richard, Phou, Fromental,
  Pomeau, and Audoly}]{Chakrabarti2018}
\bibinfo{author}{A.~Chakrabarti}, \bibinfo{author}{S.~Mora},
  \bibinfo{author}{F.~Richard}, \bibinfo{author}{T.~Phou},
  \bibinfo{author}{J.-M. Fromental}, \bibinfo{author}{Y.~Pomeau},
  \bibinfo{author}{B.~Audoly},
\newblock \bibinfo{title}{Selection of hexagonal buckling patterns by the
  elastic {R}ayleigh--{T}aylor instability},
\newblock \bibinfo{journal}{J. Mech. Phys. Solids} \bibinfo{volume}{121}
  (\bibinfo{year}{2018}) \bibinfo{pages}{234–--257}.
\bibitem[{Chiu-Webster and Lister(2006)}]{ChiuWebster2006}
\bibinfo{author}{S.~Chiu-Webster}, \bibinfo{author}{J.~R. Lister},
\newblock \bibinfo{title}{The fall of a viscous thread onto a moving surface: a
  ‘fluid-mechanical sewing machine’},
\newblock \bibinfo{journal}{J. Fluid Mech.} \bibinfo{volume}{569}
  (\bibinfo{year}{2006}) \bibinfo{pages}{89--111}.
\bibitem[{Audoly et~al.(2013)Audoly, Clauvelin, Brun, Bergou, Grinspun, and
  Wardetzky}]{Audoly2013}
\bibinfo{author}{B.~Audoly}, \bibinfo{author}{N.~Clauvelin},
  \bibinfo{author}{P.-T. Brun}, \bibinfo{author}{M.~Bergou},
  \bibinfo{author}{E.~Grinspun}, \bibinfo{author}{M.~Wardetzky},
\newblock \bibinfo{title}{A discrete geometric approach for simulating the
  dynamics of thin viscous threads},
\newblock \bibinfo{journal}{J. Comp. Phys.} \bibinfo{volume}{253}
  (\bibinfo{year}{2013}) \bibinfo{pages}{18–--49}.
\bibitem[{Jawed et~al.(2014)Jawed, Da, Joo, Grinspun, and Reis}]{Jawed2014}
\bibinfo{author}{M.~K. Jawed}, \bibinfo{author}{F.~Da},
  \bibinfo{author}{J.~Joo}, \bibinfo{author}{E.~Grinspun},
  \bibinfo{author}{P.~M. Reis},
\newblock \bibinfo{title}{Coiling of elastic rods on rigid substrates},
\newblock \bibinfo{journal}{Proc. Natl Acad. Sci. USA} \bibinfo{volume}{111}
  (\bibinfo{year}{2014}) \bibinfo{pages}{14663–--14668}.
\bibitem[{Jawed et~al.(2015)Jawed, Brun, and Reis}]{Jawed2015}
\bibinfo{author}{M.~K. Jawed}, \bibinfo{author}{P.~T. Brun},
  \bibinfo{author}{P.~M. Reis},
\newblock \bibinfo{title}{A geometric model for the coiling of an elastic rod
  deployed onto a moving substrate},
\newblock \bibinfo{journal}{J. Appl. Mech.} \bibinfo{volume}{82}
  (\bibinfo{year}{2015}) \bibinfo{pages}{121007}.
\bibitem[{Simitses(1990)}]{Simitses1990}
\bibinfo{author}{G.~J. Simitses}, \bibinfo{title}{Dynamic Stability of Suddenly
  Loaded Structures}, \bibinfo{publisher}{Springer}, \bibinfo{year}{1990}.
\bibitem[{Seffen(2006)}]{Seffen2006}
\bibinfo{author}{K.~A. Seffen},
\newblock \bibinfo{title}{Mechanical memory metal: a novel material for
  developing morphing elastic structures},
\newblock \bibinfo{journal}{Scr. Mater.} \bibinfo{volume}{55}
  (\bibinfo{year}{2006}) \bibinfo{pages}{411--–414}.
\bibitem[{Chung et~al.(2018)Chung, Vaziri, and Mahadevan}]{Chung2018}
\bibinfo{author}{J.~Y. Chung}, \bibinfo{author}{A.~Vaziri},
  \bibinfo{author}{L.~Mahadevan},
\newblock \bibinfo{title}{Reprogrammable braille on an elastic shell},
\newblock \bibinfo{journal}{Proc. Natl Acad. Sci.} \bibinfo{volume}{115}
  (\bibinfo{year}{2018}) \bibinfo{pages}{7509--–7514}.
\bibitem[{Ruhoff et~al.(1996)Ruhoff, Pr{\ae}stgaard, and
  Perram}]{ruhoff1996efficient}
\bibinfo{author}{P.~T. Ruhoff}, \bibinfo{author}{E.~Pr{\ae}stgaard},
  \bibinfo{author}{J.~W. Perram},
\newblock \bibinfo{title}{Efficient algorithms for simulating complex
  mechanical systems using constraint dynamics},
\newblock \bibinfo{journal}{Proc. R. Soc. A} \bibinfo{volume}{452}
  (\bibinfo{year}{1996}) \bibinfo{pages}{1139--1165}.
\bibitem[{Dichmann and Maddocks(1996)}]{dichmann1996impetus}
\bibinfo{author}{D.~J. Dichmann}, \bibinfo{author}{J.~H. Maddocks},
\newblock \bibinfo{title}{An impetus-striction simulation of the dynamics of an
  elastica},
\newblock \bibinfo{journal}{J. Nonlin. Sci.} \bibinfo{volume}{6}
  (\bibinfo{year}{1996}) \bibinfo{pages}{271--292}.

\end{thebibliography}

\end{document}